\documentclass[12pt,aps,prb,nofootinbib,longbibliography,superscriptaddress]{revtex4-2}
\usepackage{amsmath}
\usepackage{amssymb}
\usepackage{bm}
\usepackage{epsfig}
\usepackage{graphicx}
\usepackage{color}
\usepackage{booktabs}

\newcommand{\e}{{\rm e}}

\newcommand{\Q}{\mathcal{Q}}

\usepackage{cancel}

 \usepackage{color}
 \definecolor{blueviolet}{rgb}{0.54,0.17,0.89}
\definecolor{darkgreen}{rgb}{0, 0.5, 0}

\usepackage{soul}
\usepackage{calc}
\newsavebox\CBox\newcommand\hcancel[2][0.5pt]{%
  \ifmmode\sbox\CBox{$#2$}\else\sbox\CBox{#2}\fi%
  \makebox[0pt][l]{\usebox\CBox}%
  \rule[0.5\ht\CBox-#1/2]{\wd\CBox}{#1}} 

\def\be{\begin{equation}}
\def\ee{\end{equation}}
\def\bea{\begin{eqnarray}}
\def\eea{\end{eqnarray}}

\newcommand{\ket}[1]{\left|{#1}\right\rangle}
\newcommand{\bra}[1]{\left\langle{#1}\right|}

\makeatletter
\def\maketitle{
\@author@finish
\title@column\titleblock@produce
\suppressfloats[t]}
\makeatother

\begin{document}

\newcount\timehh  \newcount\timemm
\timehh=\time \divide\timehh by 60
\timemm=\time
\count255=\timehh\multiply\count255 by -60 \advance\timemm by \count255

\title{Extinction Coefficients of CdSe, CdS, and CdTe Nanoplatelets in Solution: A Practical Tool for Concentration Determination}

\author{Michael H. Stewart}
\affiliation{Optical Sciences Division, U.S. Naval Research Laboratory, Washington, DC 20375, USA.}
\email{michael.h.stewart30.civ@us.navy.mil; michael.w.swift5.civ@us.navy.mil; alex.l.efros.civ@us.navy.mil}
\author{Michael W. Swift}
\affiliation{Center for Computational Materials Science, U.S. Naval Research Laboratory, Washington, DC 20375, USA.}
\author{Farwa Awan}
\affiliation{Department of Chemistry, University of Rochester, Rochester, NY 14627, USA.}
\author{Liam Burke}
\affiliation{Department of Chemistry, University of Rochester, Rochester, NY 14627, USA.}
\author{Christopher M. Green}
\affiliation{Center for Biomolecular Science and Engineering, U.S. Naval Research Laboratory, Washington, DC 20375, USA.}
\author{Barbara A. Marcheschi}
\affiliation{Optical Sciences Division, U.S. Naval Research Laboratory, Washington, DC 20375, USA.}
\author{Igor L. Medintz}
\affiliation{Center for Biomolecular Science and Engineering, U.S. Naval Research Laboratory, Washington, DC 20375, USA.}
\author{Todd D. Krauss}
\affiliation{Department of Chemistry, University of Rochester, Rochester, NY 14627, USA.}
\affiliation{Institute of Optics, University of Rochester, Rochester, NY 14627, USA.}
\author{Alexander L. Efros}
\affiliation{Center for Computational Materials Science, U.S. Naval Research Laboratory, Washington, DC 20375, USA.}

\begin{abstract}
    Semiconductor nanoplatelets possess exceptional optical properties that make them promising candidates for next-generation optoelectronic applications. However, unlike quantum dots where absorption spectroscopy alone can determine both size and concentration, nanoplatelets present a significant characterization challenge: the absorption peak position reveals only thickness, providing no information about lateral dimensions or concentration. This limitation forces researchers to rely on time-consuming and costly elemental analysis techniques for complete sample characterization. Here, we present an experimentally verified theoretical framework that predicts the frequency-dependent absorption coefficient of randomly oriented CdSe, CdS, and CdTe nanoplatelets, enabling concentration determination from absorption measurements and lateral size estimates. Our model shows that the integrated absorption coefficient depends universally on nanoplatelet surface area and thickness, yielding a practical tool to extract concentrations without laborious elemental analysis. This approach bridges the characterization gap between quantum dots and nanoplatelets, offering a streamlined method for rapid sample analysis that could accelerate nanoplatelet research and applications.
\end{abstract}

\maketitle

Semiconductor nanoplatelets (NPLs) represent a unique class of two-dimensional nanocrystals that combine the advantages of colloidal synthesis with quasi-two-dimensional quantum confinement.\cite{ithurria_quasi_2008,ithurria_colloidal_2011} These atomically flat structures, typically only a few monolayers thick but tens of nanometers in lateral extent, exhibit remarkable optical properties including narrow emission linewidths, high quantum yields, and exceptionally large absorption cross-sections.\cite{shornikova_addressing_2018,meerbach_brightly_2019,shornikova_negatively_2020} The precise control over thickness at the atomic level allows for systematic tuning of emission wavelength across a broad spectral range, making NPLs attractive for applications in displays, lighting, and photovoltaic devices.

The strong quantum confinement in the thickness direction, combined with weak confinement in the lateral plane, results in enhanced Coulomb interactions and large exciton binding energies that can exceed several hundred meV.\cite{shornikova_exciton_2021,efros_nanocrystal_2021} This strong binding ensures stable photoluminescence even at room temperature, while the giant oscillator strength and reduced dielectric screening at normal incidence lead to ultrafast radiative recombination and high brightness.\cite{swift_dark_2022,swift_controlling_2024} Additionally, the unique electronic structure of NPLs enables efficient energy transfer processes, making them excellent donors and acceptors in F\"{o}rster resonance energy transfer applications~\cite{rowland_picosecond_2015,taghipour_near-unity_2018,erdem_orientation-controlled_2019}.

Despite these promising optical properties, NPLs face significant characterization challenges that limit their widespread adoption compared to spherical quantum dots. For quantum dots, the well-established relationship between absorption peak position and particle size allows researchers to determine both diameter and concentration from a single absorption spectrum.\cite{yu_experimental_2003} This convenience has made quantum dots accessible to researchers across various fields. In contrast, NPLs present a more complex characterization problem where the absorption peak position reveals only the thickness, providing no information about lateral dimensions or particle concentration. This characterization limitation forces researchers working with NPLs to rely on multiple complementary techniques. Transmission electron microscopy (TEM) becomes essential for determining lateral dimensions and size distributions, while concentration measurements typically require destructive elemental analysis techniques such as inductively coupled plasma optical emission spectroscopy (ICP-OES) or mass spectroscopy (ICP-MS).\cite{yeltik_experimental_2015} Pioneering work by Yeltik et al. provided the first systematic experimental determination of absorption cross-sections for CdSe NPLs with varying lateral sizes, establishing empirical relationships between NPL dimensions and optical properties.\cite{yeltik_experimental_2015} However, their approach still required extensive elemental analysis to validate concentration measurements, making it resource-intensive and potentially wasteful of valuable samples.  Rod\`a et al. showed the linear dependence of exciton oscillator strength on nanoplatelet area, but did not discuss the implications for concentration measurement.~\cite{roda_area-independence_2022}

The development of theoretical models that can predict absorption coefficients based solely on NPL geometry would represent a significant advancement for the field. Such models could enable researchers to determine concentrations from simple absorption measurements combined with lateral size estimates from TEM, eliminating the need for time-consuming elemental analysis.

In this work, we present a comprehensive theoretical framework for calculating the frequency-dependent absorption coefficient of randomly oriented CdSe, CdS and CdTe  NPLs. Our model demonstrates that the molar extinction coefficient depends universally on NPL surface area and thickness, independent of the specific aspect ratio. We validate our theoretical predictions against experimental data  in CdSe NPLs and provide practical guidelines for extracting NPL concentrations from routine optical measurements. This approach offers experimentalists a streamlined characterization tool that could significantly accelerate NPL research and development.


The optical properties of solutions of semiconductor nanocrystals are governed by the complex refractive index $\mathcal{N} = n - ik$, where the extinction coefficient $k$ determines the absorption characteristics.\cite{fox_optical_2008} At frequency $\omega$, the intensity $I$ of light propagating a distance $z$ is given by $I(\omega,z) = I(\omega,0) e^{-\alpha(\omega) z}$, where the absorption coefficient $\alpha(\omega)$ is related to the extinction coefficient through $k = \alpha(\omega)c/(2\omega n)$, where $c$ is the speed of light.\cite{landau_quantum_2007,fox_optical_2008}

For randomly oriented NPLs in solution, the absorption coefficient must account for the statistical average over all possible orientations relative to the incident light polarization. Absorption from an NPL is maximized when it is aligned perpendicular to the absorbing light, because the lowest exciton state arises from the heavy hole subband, which has two orthogonal optical transition dipoles lying in the NPL plane. We find that the absorption coefficient for an ensemble of randomly oriented NPLs is:
\begin{equation}
\alpha(\omega) = \frac{2}{3}N S(\omega)~,
\label{eq:random_absorption}
\end{equation}
where $N$ is the NPL concentration, $S(\omega)$ is the absorption cross-section of a single NPL oriented normal to the incident light, and the factor of $2/3$ accounts for the orientational average (see the Supporting Information for details).

The absorption cross-section $S$ of an individual NPL is defined by~\cite{fox_optical_2008}
\begin{equation}
I(\omega) \frac{c}{n} S(\omega)=\hbar\omega  P(\omega)~,
 \label{eq:S_def}
 \end{equation}
 where $P(\omega)$   is  total excitation probability of a single particle by the light.  Therefore $S(\omega)$ can be calculated using the transition dipole moment and the density of states. For CdSe, CdS and CdTe NPLs, the dominant optical transition involves heavy-hole to conduction band excitons with a transition energy determined by the thickness-dependent quantum confinement.  It can be shown (see the Supporting Information for details) that the cross-section takes the form:
\begin{equation}
S(\omega) = {2\hbar^2\pi nE_\text{p}\over  137 m_0 E_\text{ex} a_\text{2D}^2}\Q f_T \sum_i\left|\int d^2R\Psi_i(\bm R)\right|^2\delta(E_i-\hbar\omega)~.
\label{eq:S_omega}
\end{equation}
In this equation, $E_p$ is the Kane energy parameter, $m_0$ is the electron mass, $E_\text{ex}$ is the exciton transition energy, $a_\text{2D}$ is the exciton radius, and $\Psi_i(\bm R)$ are the exciton center-of-mass envelope wavefunctions with corresponding to energy  $E_i$ of the 2D exciton  motion in the NPL plane.  The factor $\Q$ takes into account non-parabolicity of the electron conduction bands. The factor $f_T$ accounts for the temperature-dependent dephasing that broadens the exciton linewidth and reduces the peak absorption coefficient, which we describe phenomenologically through, \cite{efros_electronic_2000,swift_controlling_2024}
\begin{equation}
  f_T = (1-\gamma_\text{ph} e^{-\hbar \omega_\text{ph}/k_\text{B} T})~.
  \label{eq:f_T}
\end{equation}
Here $\omega_\text{ph} = \pi v_\text{ph} /d$ is the energy of acoustic phonons  that are  confined in the NPL  with  thickness $d$,  and  $v_\text{ph}$ is the phonon velocity.  The dimensionless parameter $\gamma_\text{ph}$ describes the coupling between the ground-state exciton and the phonons.  It is considered to be between 0 and 1 and we use it as a fitting parameter.

Completeness of the envelope functions $\Psi_i$ allows a dramatic simplification if we integrate over frequencies:~\cite{swift_controlling_2024}
\begin{equation}
  \int \hbar\, d\omega  \sum_i\left|\int d^2R\Psi_i(\bm R)\right|^2\delta(E_i-\hbar\omega) = S_\text{NPL}~,
\end{equation}
where $S_\text{NPL}$ is the lateral surface area of the NPL.  Therefore, integrating the absorption coefficient over the frequency range of the first exciton peak, we obtain the key result:
\begin{equation}
\overline{\alpha}=\int  \hbar d\omega \alpha(\hbar \omega)=N S_\text{NPL} \frac{4\pi \hbar^2 nE_\text{p} }{411m_0} \frac{\Q f_T}{E_\text{ex} a_\text{2D}^2} ~.
\label{eq:22}
\end{equation}
This expression reveals that the integrated absorption coefficient depends only on the NPL concentration, lateral area, and thickness (through  $\Q$, $f_T$, $a_\text{2D}$, and $E_\text{ex}$). This universal dependence forms the basis for our practical concentration determination method.

We can now make connections to experimentally accessible observables. The Beer-Lambert law states that $A= b N E$, where $A = \log_{10}(I_0/I)$ is the absorbance, $b$ is the path length, $N$ is the concentration, and $E$ is the molar extinction coefficient.  Therefore, $E = \alpha/(N\ln(10))$, where $N$ is the nanoparticle concentration.  The integrated molar absorption coefficient is obtained by integrating $E$ over the first exciton peak, with energy as the horizontal axis:
\begin{align}
  \overline{E} = \int \hbar \,d\omega E(\omega) =  S_\text{NPL} \frac{4\pi \hbar^2 nE_\text{p} }{411\ln(10)m_0} \frac{\Q f_T}{E_\text{ex} a_\text{2D}^2} \approx 0.0133 S_\text{NPL} \frac{\hbar^2 nE_\text{p} }{m_0} \frac{\Q f_T}{E_\text{ex} a_\text{2D}^2} 
  \label{eq:overline_E}
\end{align}
Assuming the absorption peak is Gaussian with full width at half maximum $\Delta$, the peak value is
\begin{equation}
  E_\text{max} = 2\sqrt{\frac{\ln (2)}{\pi}}\, \frac{\overline{E}}{ \Delta} \approx 0.0125 S_\text{NPL} \frac{\hbar^2 nE_\text{p} }{m_0} \frac{\Q f_T}{ E_\text{ex} a_\text{2D}^2 \Delta} 
  \label{eq:Emax}
\end{equation}
The calculation of $\overline E$ is more accurate, since it does not assume a Gaussian absorption line.  However, $E_\text{max}$ is a more easily accessible experimental quantity, making this a valuable alternate point of comparison.

\begin{figure}
\includegraphics[width=\textwidth]{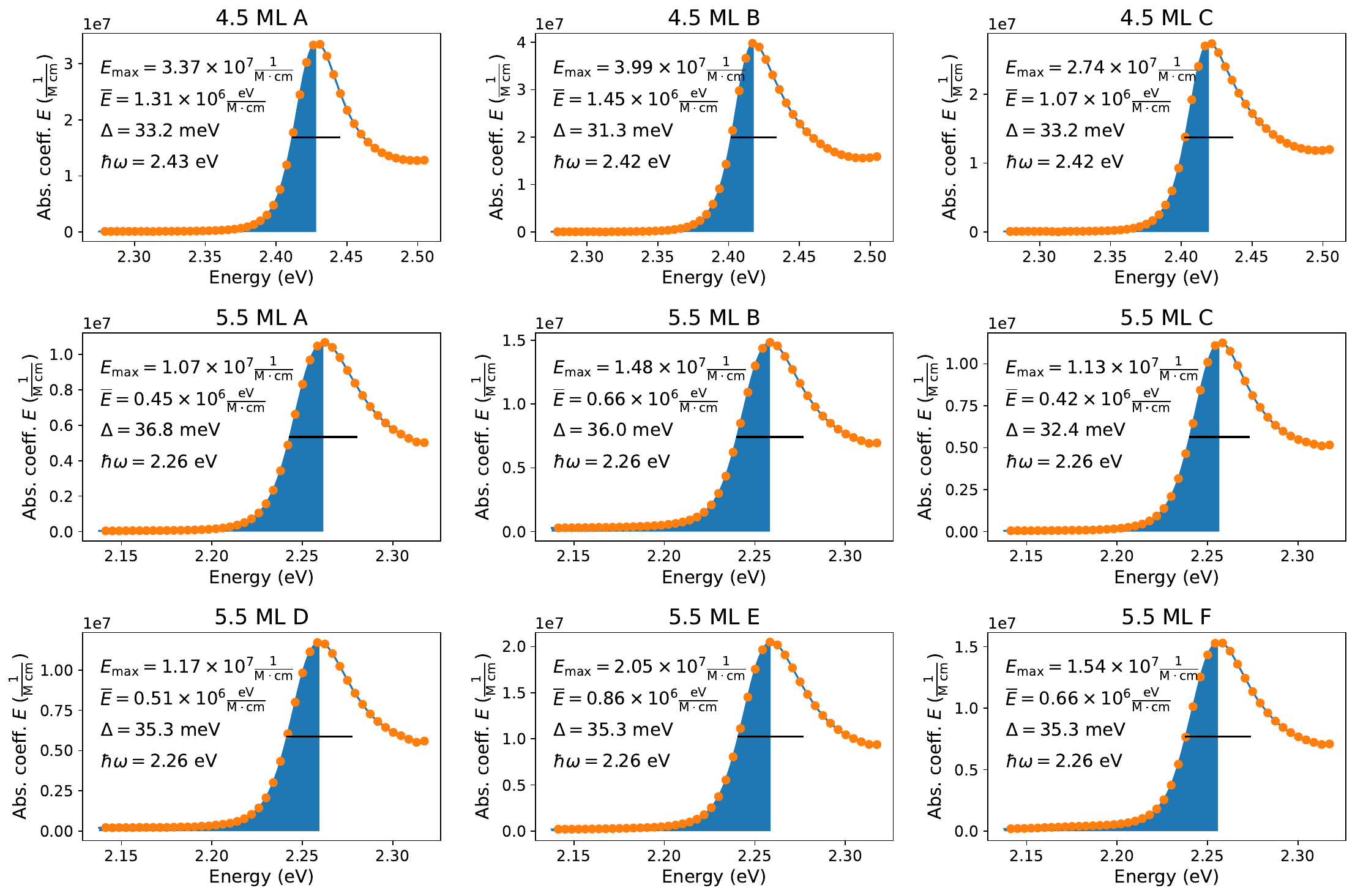}
\caption{Experimental absorption spectra (orange) with cubic spline interpolation (blue) for CdSe NPLs of varying thickness and lateral size. The integrated absorption coefficient $\overline{E}$ is calculated from the area under the first exciton peak (blue shaded region). The linewidth parameter $\Delta$, shown as a black horizontal line, is determined by doubling the left half-width at half-maximum. 4.5 ML samples are labeled A-C, and 5.5 ML samples are labeled A-F.}
\label{fig:Peak_integration}
\end{figure}

For experimental comparison, we chose CdSe NPLs (4.5 ML and 5.5 ML) as an ideal system due to their reliable synthesis protocols that provide pure material with tunable lateral areas. Figure 1 shows experimental absorption spectra for NPLs of different thicknesses and lateral sizes. The CdSe NPLs were synthesized with slight modification of published procedures to control the lateral sizes.\cite{bertrand_shape_2016,dutta_hybrid_2020,khan_tunable_2019}  The lateral areas for the NPL samples are shown in Table 1, which range from 264-329 nm$^2$ and 173-342 nm$^2$ for 4.5 ML and 5.5 ML CdSe NPLs, respectively. The dimensions of each NPL sample were estimated from analysis of the TEM images. NPL concentrations were determined by measuring the Se concentration of acid digested NPL samples using inductively coupled plasma optical emission spectroscopy (ICP-OES)~\cite{,morrison_methods_2020} and applying an analysis method established by Yeltik et al.\cite{yeltik_experimental_2015}. The samples were rigorously purified to remove any impurities that might interfere with the subsequent analytical analysis. Using the Beer-Lambert law, we could extract experimental extinction coefficients for the NPLs show in Figure 2.  To calculate $\overline{E}$, the molar extinction coefficient data is interpolated by a cubic spline and integrated up to the first peak, and the result is multiplied by 2, as illustrated in Figure~\ref{fig:Peak_integration}.  Similarly, in order to calculate the full-width at half-maximum $\Delta$, the left half-width is measured and multiplied by 2.

The practical implementation of our model requires knowledge of several material parameters for CdSe: the Kane energy parameter ($E_p \approx 17.5$ eV), the refractive index ($n \approx 1.5$), and the thickness-dependent exciton binding parameters. For 4.5 ML NPLs, we use an exciton transition energy of $\hbar\omega = 2.42$ eV and exciton radius $a_{2D} = 1.85$ nm. For 5.5 ML NPLs, the corresponding values are $\hbar\omega = 2.26$ eV and $a_{2D} = 2.05$ nm.  These parameters can be determined from independent measurements or calculated using effective mass theory~\cite{swift_controlling_2024}. For the phonon velocity in Eq.~\ref{eq:f_T}, we use the transverse acoustic phonon $v_\text{ph} = 1.90 \times 10^5$ cm/s~\cite{hieu_temperature-dependent_2024}.  All measurements and calculations are done at 300 K.  

\begin{figure}
  \includegraphics[width=0.49\textwidth]{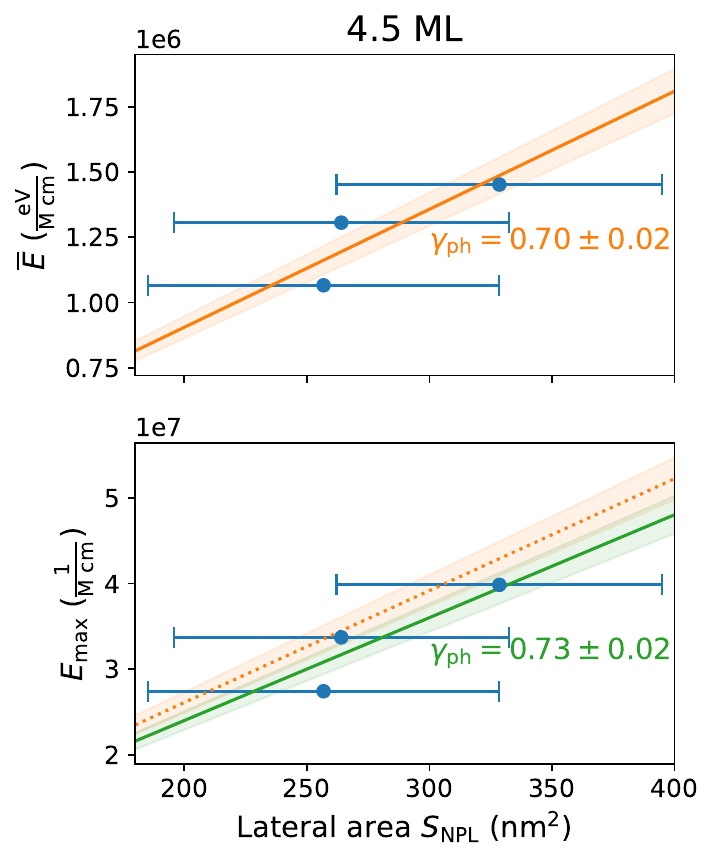}
  \includegraphics[width=0.49\textwidth]{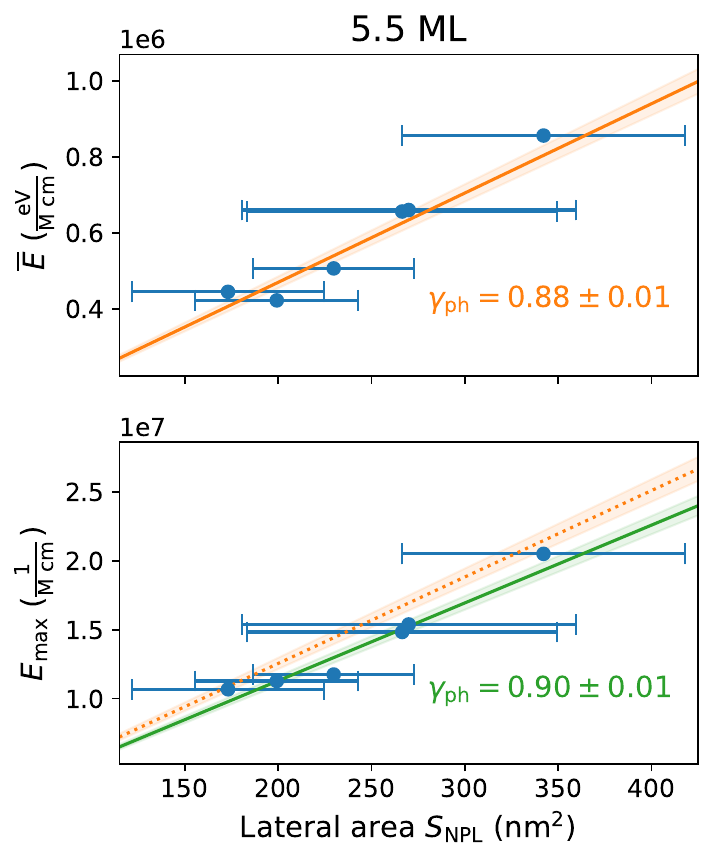}
  \caption{Integrated extinction coefficient $\overline{E}$ (top) and peak extinction coefficient $E_\text{max}$ (bottom) as a function of lateral area in CdSe NPLs with thickness 4.5 ML (left) and 5.5 ML (right).  Blue dots show experimental data.  Orange lines fit Eq.~(\ref{eq:overline_E}) to the $\overline{E}$ data, with the fitted $\gamma_\text{ph}$ shown in orange on the top panels.  The slopes of the orange lines are $\overline{E}/S_\text{NPL} =  4524 \frac{\text{eV}}{\text{M cm nm}^2}$ for 4.5 ML NPLs and  $2350 \frac{\text{eV}}{\text{M cm nm}^2}$ for 5.5 ML NPLs.  The green line fits Eq.~(\ref{eq:Emax}) to the $E_\text{max}$ data, with the fitted $\gamma_\text{ph}$ shown in green on the bottom panels.}
  \label{fig:E_comparison}
\end{figure}

When comparing theory with experiment (Fig.~\ref{fig:E_comparison}), we find excellent agreement. $\overline{E}$ is proportional to the nanoplatelet area, with slopes $\overline{E}/S_\text{NPL} = 4524 \frac{\text{eV}}{\text{M cm nm}^2}$ for 4.5 ML NPLs and  $2350 \frac{\text{eV}}{\text{M cm nm}^2}$ for 5.5 ML NPLs.  The phonon coupling parameter $\gamma_\text{ph}$ (see Eq.~\ref{eq:f_T}) is the only adjustable parameter in our model, with fitted values of approximately 0.7-0.9 for both 4.5 ML and 5.5 ML NPLs.  

Our theoretical framework provides experimentalists with a straightforward protocol for concentration determination, eliminating the need for complicated elemental analysis. First, the absorbance is measured and integrated over the first exciton peak with respect to energy, obtaining $\overline{A}$.  Second, TEM analysis provides the lateral area $S_\text{NPL}$. Third, the averaged molar extinction coefficient $\overline{E}$ is obtained.  This may be extracted from Figure~\ref{fig:E_comparison}, calculated by multiplying $S_\text{NPL}$ by the calculated slopes, or calculated directly using Eq.~\ref{eq:Emax}.  Finally, the concentration is given by $N = \overline{A}/b\overline{E}$.  An example is shown in Table~\ref{tab:experimental_data}, in which the concentration calculated using this procedure may be compared with the concentration measured by ICP-OES.  

\begin{table}
\caption{Experimental data for CdSe NPLs at 300K. Thickness $d$ is given in monolayers (ML), and samples are labled as in Figure~\ref{fig:Peak_integration}. The lateral area $S_\text{NPL}$ is reported as mean $\pm$ standard deviation from TEM analysis, absorbance is integrated with respect to energy up to the first exciton peak and doubled to obtain $\overline{A}$, experimental concentration $N$ is determined from ICP-OES measurements, theoretical $\overline{E}$ is from the orange lines in Figure~\ref{fig:E_comparison}, and predicted concentration $N$ is calculated using $N = \overline{A}/b\overline{E}$, where the path length $b$ is 1 cm.}
\label{tab:experimental_data}
\begin{tabular}{cccccc}
\toprule
Label & $S_\text{NPL}$  & $\overline{A}$ (expt)  & $N$ (expt)  & $\overline{E}$ (theor) & Predicted $N$ \\
 & [nm$^2$] & [meV] & [nM] & [meV/(nM$\cdot$cm)] & [nM]  \\
\midrule
4.5 ML A & 264 $\pm$ 68 & 33.44 & 25.6 & 1.195 & 28.0 \\
4.5 ML B & 329 $\pm$ 66 & 20.77 & 14.3 & 1.487 & 14.0 \\
4.5 ML C & 257 $\pm$ 72 & 12.47 & 11.7 & 1.162 & 10.7 \\
5.5 ML A & 173 $\pm$ 52 & 7.49 & 16.8 & 0.407 & 18.4 \\
5.5 ML B & 266 $\pm$ 83 & 6.15 & 9.4 & 0.626 & 9.8 \\
5.5 ML C & 199 $\pm$ 44 & 5.92 & 14.0 & 0.468 & 12.6 \\
5.5 ML D & 230 $\pm$ 43 & 3.95 & 7.8 & 0.540 & 7.3 \\
5.5 ML E & 342 $\pm$ 76 & 8.40 & 9.8 & 0.804 & 10.4 \\
5.5 ML F & 270 $\pm$ 89 & 7.39 & 11.2 & 0.634 & 11.6 \\
\bottomrule
\end{tabular}
\end{table}

\begin{figure}
  \includegraphics[width=\textwidth]{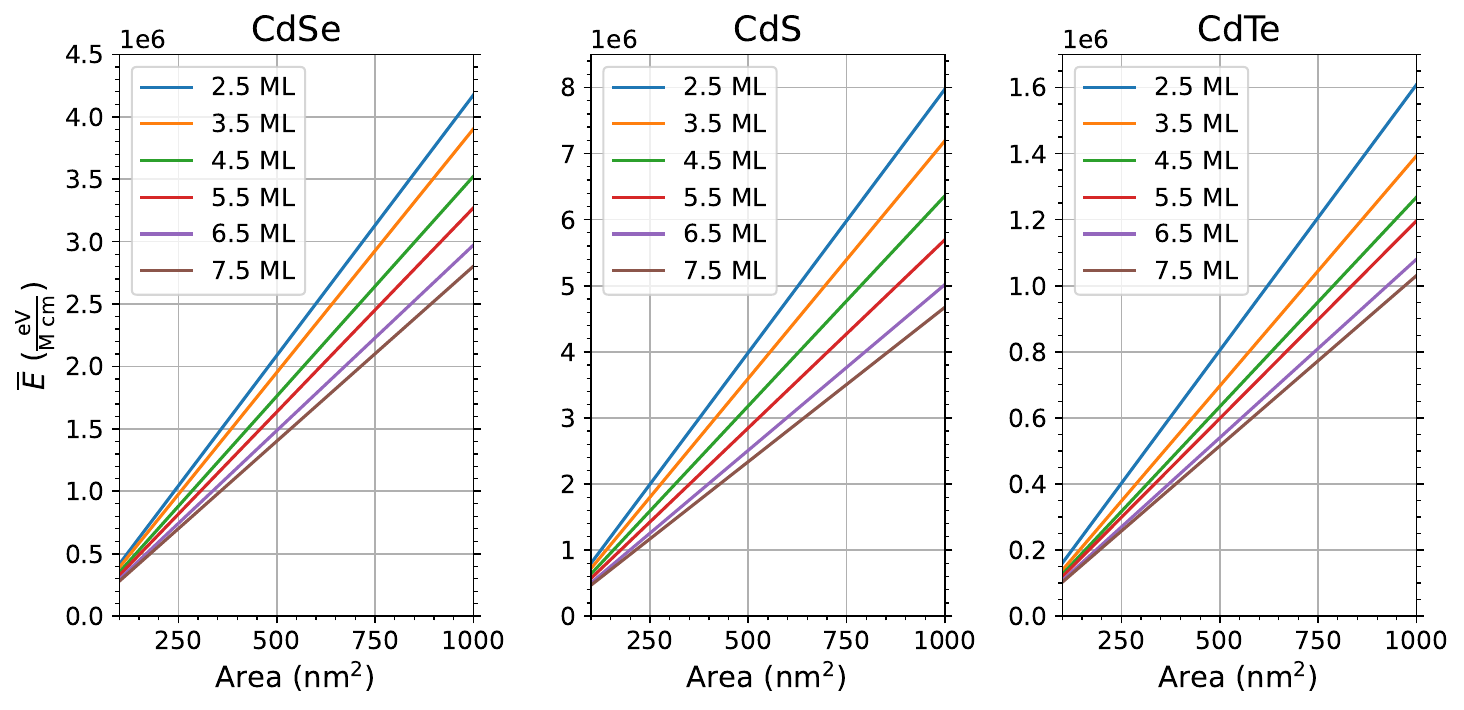}
  \caption{Integrated extinction coefficient, $\overline{E}$, plotted as a function of temperature in CdSe, CdS, and CdTe NPLs with thickness 2.5 through 7.5 ML as shown on the legends.  Phonon coupling factor $\gamma_\text{ph} = 0.79$ is assumed.}
  \label{fig:E_prediction}
\end{figure}

We can generalize to other thicknesses by assuming $\gamma_\text{ph} = 0.79$, the average value across 4.5 ML and 5.5 ML NPLs.  If we assume this $\gamma_\text{ph}$ holds across CdSe and CdTe as well, we can use material parameters from the literature~\cite{ninomiya_optical_1995,marple_refractive_1964} (See SI Table~\ref{s_tab:material}) to calculate $\overline{E}$ for CdSe and CdTe using Eq.~\ref{eq:overline_E}.  The result is shown in Figure~\ref{fig:E_prediction}, and the slopes of the plotted lines are given in Table~\ref{tab:E_slope_prediction}.  These results extend the protocol for concentration determination to 2.5 through 7.5 monolayer NPLs in all three materials.

\begin{table}
\caption{Predicted slopes $\overline{E}/S_\text{NPL}$ (in units of $\frac{\text{eV}}{\text{M cm nm}^2}$) for CdSe, CdS, and CdTe NPLs as a function of thickness, assuming phonon coupling factor $\gamma_\text{ph} = 0.79$. }
\label{tab:E_slope_prediction}
\begin{tabular}{lrrr}
  \toprule
 & CdSe & CdS & CdTe \\
 \midrule
2.5 ML & 4222 & 8051 & 1629 \\
3.5 ML & 3884 & 7193 & 1381 \\
4.5 ML & 3586 & 6288 & 1268 \\
5.5 ML & 3191 & 5729 & 1189 \\
6.5 ML & 2966 & 5225 & 1089 \\
7.5 ML & 2756 & 4773 & 1057 \\
\bottomrule
\end{tabular}
\end{table}


We have developed a comprehensive theoretical framework for calculating the absorption coefficient of randomly oriented CdSe, CdS and CdTe  NPLs that addresses a critical characterization challenge in the field. Unlike quantum dots, where absorption spectroscopy alone can determine both size and concentration, NPLs have historically required multiple complementary techniques including TEM for lateral dimensions and elemental analysis for concentration determination. Our model bridges this gap by providing a direct relationship between easily measurable quantities—lateral area from TEM and integrated absorption from UV-visible spectroscopy—and NPL concentration.

The key theoretical insight is that the frequency-integrated absorption coefficient depends universally on NPL surface area and thickness, independent of specific aspect ratios or detailed geometry. This universal scaling enables practical concentration determination without requiring time-consuming elemental analysis techniques such as ICP-OES. The theoretical predictions show excellent agreement with experimental data when phonon-induced decoherence effects are included, validating the model's accuracy and practical utility.

For the experimental community, this work provides a streamlined characterization protocol that could significantly accelerate NPL research. By combining routine TEM analysis with absorption spectroscopy, researchers can now determine all essential sample parameters—thickness, lateral dimensions, and concentration—without specialized analytical equipment or sample destruction. This accessibility should lower barriers to NPL research and enable broader adoption of these promising nanomaterials.

The theoretical framework presented here establishes the foundation for extending similar approaches to other NPL materials and geometries. As synthetic methods continue to advance and enable new NPL compositions and structures, having robust theoretical tools for characterization will become increasingly important. Our work demonstrates that fundamental optical theory can provide practical solutions to experimental challenges, bridging the gap between theoretical understanding and laboratory practice.

Looking forward, this characterization methodology should facilitate more systematic studies of NPL properties and applications. With simplified sample analysis, researchers can focus on optimizing synthetic conditions, exploring new applications, and developing NPL-based devices. The combination of exceptional optical properties and accessible characterization methods positions NPLs as increasingly attractive alternatives to quantum dots for next-generation optoelectronic applications.

\section*{Acknowledgments}
The authors acknowledge Alexander Searle, Sreeparna Banerjee, and  Lisbeth Crompton  for their help with sample characterization and Peter Sercel  for his comments on the manuscript.  M.~H.~S., M.~W.~S., C.~M.~G., B.~A.~M., I.~L.~M., and Al.~L.~E. acknowledge the support of the Office of Naval Research. NIPRGPT with Anthropic Claude 4 Sonnet was used to edit the manuscript text.  The synthesis and characterization of the NPLs was supported by the National Science Foundation (NSF) under Grant No. CHE-2304937 .

\section*{Author information}
Michael H Stewart https://orcid.org/0000-0003-1415-1590

Michael W Swift https://orcid.org/0000-0003-2974-6052

Farwa Awan https://orcid.org/0000-0002-4973-0966

Liam Burke https://orcid.org/0009-0003-7216-7977

Todd D. Krauss https://orcid.org/0000-0002-4860-874X

Alexander L. Efros https://orcid.org/0000-0003-1938-553X


\begin{thebibliography}{25}%
\makeatletter
\providecommand \@ifxundefined [1]{%
 \@ifx{#1\undefined}
}%
\providecommand \@ifnum [1]{%
 \ifnum #1\expandafter \@firstoftwo
 \else \expandafter \@secondoftwo
 \fi
}%
\providecommand \@ifx [1]{%
 \ifx #1\expandafter \@firstoftwo
 \else \expandafter \@secondoftwo
 \fi
}%
\providecommand \natexlab [1]{#1}%
\providecommand \enquote  [1]{``#1''}%
\providecommand \bibnamefont  [1]{#1}%
\providecommand \bibfnamefont [1]{#1}%
\providecommand \citenamefont [1]{#1}%
\providecommand \href@noop [0]{\@secondoftwo}%
\providecommand \href [0]{\begingroup \@sanitize@url \@href}%
\providecommand \@href[1]{\@@startlink{#1}\@@href}%
\providecommand \@@href[1]{\endgroup#1\@@endlink}%
\providecommand \@sanitize@url [0]{\catcode `\\12\catcode `\$12\catcode `\&12\catcode `\#12\catcode `\^12\catcode `\_12\catcode `\%12\relax}%
\providecommand \@@startlink[1]{}%
\providecommand \@@endlink[0]{}%
\providecommand \url  [0]{\begingroup\@sanitize@url \@url }%
\providecommand \@url [1]{\endgroup\@href {#1}{\urlprefix }}%
\providecommand \urlprefix  [0]{URL }%
\providecommand \Eprint [0]{\href }%
\providecommand \doibase [0]{https://doi.org/}%
\providecommand \selectlanguage [0]{\@gobble}%
\providecommand \bibinfo  [0]{\@secondoftwo}%
\providecommand \bibfield  [0]{\@secondoftwo}%
\providecommand \translation [1]{[#1]}%
\providecommand \BibitemOpen [0]{}%
\providecommand \bibitemStop [0]{}%
\providecommand \bibitemNoStop [0]{.\EOS\space}%
\providecommand \EOS [0]{\spacefactor3000\relax}%
\providecommand \BibitemShut  [1]{\csname bibitem#1\endcsname}%
\let\auto@bib@innerbib\@empty
\bibitem [{\citenamefont {Ithurria}\ and\ \citenamefont {Dubertret}(2008)}]{ithurria_quasi_2008}%
  \BibitemOpen
  \bibfield  {author} {\bibinfo {author} {\bibfnamefont {S.}~\bibnamefont {Ithurria}}\ and\ \bibinfo {author} {\bibfnamefont {B.}~\bibnamefont {Dubertret}},\ }\bibfield  {title} {\bibinfo {title} {Quasi {2D} {Colloidal} {CdSe} {Platelets} with {Thicknesses} {Controlled} at the {Atomic} {Level}},\ }\href {https://doi.org/10.1021/ja807724e} {\bibfield  {journal} {\bibinfo  {journal} {J. Am. Chem. Soc.}\ }\textbf {\bibinfo {volume} {130}},\ \bibinfo {pages} {16504} (\bibinfo {year} {2008})}\BibitemShut {NoStop}%
\bibitem [{\citenamefont {Ithurria}\ \emph {et~al.}(2011)\citenamefont {Ithurria}, \citenamefont {Tessier}, \citenamefont {Mahler}, \citenamefont {Lobo}, \citenamefont {Dubertret},\ and\ \citenamefont {Efros}}]{ithurria_colloidal_2011}%
  \BibitemOpen
  \bibfield  {author} {\bibinfo {author} {\bibfnamefont {S.}~\bibnamefont {Ithurria}}, \bibinfo {author} {\bibfnamefont {M.~D.}\ \bibnamefont {Tessier}}, \bibinfo {author} {\bibfnamefont {B.}~\bibnamefont {Mahler}}, \bibinfo {author} {\bibfnamefont {R.~P. S.~M.}\ \bibnamefont {Lobo}}, \bibinfo {author} {\bibfnamefont {B.}~\bibnamefont {Dubertret}},\ and\ \bibinfo {author} {\bibfnamefont {A.~L.}\ \bibnamefont {Efros}},\ }\bibfield  {title} {\bibinfo {title} {Colloidal nanoplatelets with two-dimensional electronic structure},\ }\href {https://doi.org/10.1038/nmat3145} {\bibfield  {journal} {\bibinfo  {journal} {Nat. Mater}\ }\textbf {\bibinfo {volume} {10}},\ \bibinfo {pages} {936} (\bibinfo {year} {2011})}\BibitemShut {NoStop}%
\bibitem [{\citenamefont {Shornikova}\ \emph {et~al.}(2018)\citenamefont {Shornikova}, \citenamefont {Biadala}, \citenamefont {Yakovlev}, \citenamefont {Sapega}, \citenamefont {Kusrayev}, \citenamefont {Mitioglu}, \citenamefont {Ballottin}, \citenamefont {Christianen}, \citenamefont {Belykh}, \citenamefont {Kochiev}, \citenamefont {Sibeldin}, \citenamefont {Golovatenko}, \citenamefont {Rodina}, \citenamefont {Gippius}, \citenamefont {Kuntzmann}, \citenamefont {Jiang}, \citenamefont {Nasilowski}, \citenamefont {Dubertret},\ and\ \citenamefont {Bayer}}]{shornikova_addressing_2018}%
  \BibitemOpen
  \bibfield  {author} {\bibinfo {author} {\bibfnamefont {E.~V.}\ \bibnamefont {Shornikova}}, \bibinfo {author} {\bibfnamefont {L.}~\bibnamefont {Biadala}}, \bibinfo {author} {\bibfnamefont {D.~R.}\ \bibnamefont {Yakovlev}}, \bibinfo {author} {\bibfnamefont {V.~F.}\ \bibnamefont {Sapega}}, \bibinfo {author} {\bibfnamefont {Y.~G.}\ \bibnamefont {Kusrayev}}, \bibinfo {author} {\bibfnamefont {A.~A.}\ \bibnamefont {Mitioglu}}, \bibinfo {author} {\bibfnamefont {M.~V.}\ \bibnamefont {Ballottin}}, \bibinfo {author} {\bibfnamefont {P.~C.~M.}\ \bibnamefont {Christianen}}, \bibinfo {author} {\bibfnamefont {V.~V.}\ \bibnamefont {Belykh}}, \bibinfo {author} {\bibfnamefont {M.~V.}\ \bibnamefont {Kochiev}}, \bibinfo {author} {\bibfnamefont {N.~N.}\ \bibnamefont {Sibeldin}}, \bibinfo {author} {\bibfnamefont {A.~A.}\ \bibnamefont {Golovatenko}}, \bibinfo {author} {\bibfnamefont {A.~V.}\ \bibnamefont {Rodina}}, \bibinfo {author} {\bibfnamefont {N.~A.}\ \bibnamefont {Gippius}}, \bibinfo {author} {\bibfnamefont {A.}~\bibnamefont {Kuntzmann}}, \bibinfo {author} {\bibfnamefont {Y.}~\bibnamefont {Jiang}}, \bibinfo {author} {\bibfnamefont {M.}~\bibnamefont {Nasilowski}}, \bibinfo {author} {\bibfnamefont {B.}~\bibnamefont {Dubertret}},\ and\ \bibinfo {author} {\bibfnamefont {M.}~\bibnamefont {Bayer}},\ }\bibfield  {title} {\bibinfo {title} {Addressing the exciton fine structure in colloidal nanocrystals: the case of {CdSe} nanoplatelets},\ }\href {https://doi.org/10.1039/C7NR07206F} {\bibfield  {journal} {\bibinfo  {journal} {Nanoscale}\ }\textbf {\bibinfo {volume} {10}},\ \bibinfo {pages} {646} (\bibinfo {year} {2018})}\BibitemShut {NoStop}%
\bibitem [{\citenamefont {Meerbach}\ \emph {et~al.}(2019)\citenamefont {Meerbach}, \citenamefont {Tietze}, \citenamefont {Voigt}, \citenamefont {Sayevich}, \citenamefont {Dzhagan}, \citenamefont {Erwin}, \citenamefont {Dang}, \citenamefont {Selyshchev}, \citenamefont {Schneider}, \citenamefont {Zahn}, \citenamefont {Lesnyak},\ and\ \citenamefont {Eychmüller}}]{meerbach_brightly_2019}%
  \BibitemOpen
  \bibfield  {author} {\bibinfo {author} {\bibfnamefont {C.}~\bibnamefont {Meerbach}}, \bibinfo {author} {\bibfnamefont {R.}~\bibnamefont {Tietze}}, \bibinfo {author} {\bibfnamefont {S.}~\bibnamefont {Voigt}}, \bibinfo {author} {\bibfnamefont {V.}~\bibnamefont {Sayevich}}, \bibinfo {author} {\bibfnamefont {V.~M.}\ \bibnamefont {Dzhagan}}, \bibinfo {author} {\bibfnamefont {S.~C.}\ \bibnamefont {Erwin}}, \bibinfo {author} {\bibfnamefont {Z.}~\bibnamefont {Dang}}, \bibinfo {author} {\bibfnamefont {O.}~\bibnamefont {Selyshchev}}, \bibinfo {author} {\bibfnamefont {K.}~\bibnamefont {Schneider}}, \bibinfo {author} {\bibfnamefont {D.~R.~T.}\ \bibnamefont {Zahn}}, \bibinfo {author} {\bibfnamefont {V.}~\bibnamefont {Lesnyak}},\ and\ \bibinfo {author} {\bibfnamefont {A.}~\bibnamefont {Eychmüller}},\ }\bibfield  {title} {\bibinfo {title} {Brightly {Luminescent} {Core}/{Shell} {Nanoplatelets} with {Continuously} {Tunable} {Optical} {Properties}},\ }\href {https://doi.org/10.1002/adom.201801478} {\bibfield  {journal} {\bibinfo  {journal} {Advanced Optical Materials}\ }\textbf {\bibinfo {volume} {7}},\ \bibinfo {pages} {1801478} (\bibinfo {year} {2019})}\BibitemShut {NoStop}%
\bibitem [{\citenamefont {Shornikova}\ \emph {et~al.}(2020)\citenamefont {Shornikova}, \citenamefont {Yakovlev}, \citenamefont {Biadala}, \citenamefont {Crooker}, \citenamefont {Belykh}, \citenamefont {Kochiev}, \citenamefont {Kuntzmann}, \citenamefont {Nasilowski}, \citenamefont {Dubertret},\ and\ \citenamefont {Bayer}}]{shornikova_negatively_2020}%
  \BibitemOpen
  \bibfield  {author} {\bibinfo {author} {\bibfnamefont {E.~V.}\ \bibnamefont {Shornikova}}, \bibinfo {author} {\bibfnamefont {D.~R.}\ \bibnamefont {Yakovlev}}, \bibinfo {author} {\bibfnamefont {L.}~\bibnamefont {Biadala}}, \bibinfo {author} {\bibfnamefont {S.~A.}\ \bibnamefont {Crooker}}, \bibinfo {author} {\bibfnamefont {V.~V.}\ \bibnamefont {Belykh}}, \bibinfo {author} {\bibfnamefont {M.~V.}\ \bibnamefont {Kochiev}}, \bibinfo {author} {\bibfnamefont {A.}~\bibnamefont {Kuntzmann}}, \bibinfo {author} {\bibfnamefont {M.}~\bibnamefont {Nasilowski}}, \bibinfo {author} {\bibfnamefont {B.}~\bibnamefont {Dubertret}},\ and\ \bibinfo {author} {\bibfnamefont {M.}~\bibnamefont {Bayer}},\ }\bibfield  {title} {\bibinfo {title} {Negatively {Charged} {Excitons} in {CdSe} {Nanoplatelets}},\ }\href {https://doi.org/10.1021/acs.nanolett.9b04907} {\bibfield  {journal} {\bibinfo  {journal} {Nano Lett.}\ }\textbf {\bibinfo {volume} {20}},\ \bibinfo {pages} {1370} (\bibinfo {year} {2020})}\BibitemShut {NoStop}%
\bibitem [{\citenamefont {Shornikova}\ \emph {et~al.}(2021)\citenamefont {Shornikova}, \citenamefont {Yakovlev}, \citenamefont {Gippius}, \citenamefont {Qiang}, \citenamefont {Dubertret}, \citenamefont {Khan}, \citenamefont {Di~Giacomo}, \citenamefont {Moreels},\ and\ \citenamefont {Bayer}}]{shornikova_exciton_2021}%
  \BibitemOpen
  \bibfield  {author} {\bibinfo {author} {\bibfnamefont {E.~V.}\ \bibnamefont {Shornikova}}, \bibinfo {author} {\bibfnamefont {D.~R.}\ \bibnamefont {Yakovlev}}, \bibinfo {author} {\bibfnamefont {N.~A.}\ \bibnamefont {Gippius}}, \bibinfo {author} {\bibfnamefont {G.}~\bibnamefont {Qiang}}, \bibinfo {author} {\bibfnamefont {B.}~\bibnamefont {Dubertret}}, \bibinfo {author} {\bibfnamefont {A.~H.}\ \bibnamefont {Khan}}, \bibinfo {author} {\bibfnamefont {A.}~\bibnamefont {Di~Giacomo}}, \bibinfo {author} {\bibfnamefont {I.}~\bibnamefont {Moreels}},\ and\ \bibinfo {author} {\bibfnamefont {M.}~\bibnamefont {Bayer}},\ }\bibfield  {title} {\bibinfo {title} {Exciton {Binding} {Energy} in {CdSe} {Nanoplatelets} {Measured} by {One}- and {Two}-{Photon} {Absorption}},\ }\href {https://doi.org/10.1021/acs.nanolett.1c04159} {\bibfield  {journal} {\bibinfo  {journal} {Nano Lett.}\ }\textbf {\bibinfo {volume} {21}},\ \bibinfo {pages} {10525} (\bibinfo {year} {2021})}\BibitemShut {NoStop}%
\bibitem [{\citenamefont {Efros}\ and\ \citenamefont {Brus}(2021)}]{efros_nanocrystal_2021}%
  \BibitemOpen
  \bibfield  {author} {\bibinfo {author} {\bibfnamefont {A.~L.}\ \bibnamefont {Efros}}\ and\ \bibinfo {author} {\bibfnamefont {L.~E.}\ \bibnamefont {Brus}},\ }\bibfield  {title} {\bibinfo {title} {Nanocrystal {Quantum} {Dots}: {From} {Discovery} to {Modern} {Development}},\ }\href {https://doi.org/10.1021/acsnano.1c01399} {\bibfield  {journal} {\bibinfo  {journal} {ACS Nano}\ }\textbf {\bibinfo {volume} {15}},\ \bibinfo {pages} {6192} (\bibinfo {year} {2021})}\BibitemShut {NoStop}%
\bibitem [{\citenamefont {Gramlich}\ \emph {et~al.}(2022)\citenamefont {Gramlich}, \citenamefont {Swift}, \citenamefont {Lampe}, \citenamefont {Lyons}, \citenamefont {Döblinger}, \citenamefont {Efros}, \citenamefont {Sercel},\ and\ \citenamefont {Urban}}]{swift_dark_2022}%
  \BibitemOpen
  \bibfield  {author} {\bibinfo {author} {\bibfnamefont {M.}~\bibnamefont {Gramlich}}, \bibinfo {author} {\bibfnamefont {M.~W.}\ \bibnamefont {Swift}}, \bibinfo {author} {\bibfnamefont {C.}~\bibnamefont {Lampe}}, \bibinfo {author} {\bibfnamefont {J.~L.}\ \bibnamefont {Lyons}}, \bibinfo {author} {\bibfnamefont {M.}~\bibnamefont {Döblinger}}, \bibinfo {author} {\bibfnamefont {A.~L.}\ \bibnamefont {Efros}}, \bibinfo {author} {\bibfnamefont {P.~C.}\ \bibnamefont {Sercel}},\ and\ \bibinfo {author} {\bibfnamefont {A.~S.}\ \bibnamefont {Urban}},\ }\bibfield  {title} {\bibinfo {title} {Dark and {Bright} {Excitons} in {Halide} {Perovskite} {Nanoplatelets}},\ }\href {https://doi.org/10.1002/advs.202103013} {\bibfield  {journal} {\bibinfo  {journal} {Advanced Science}\ }\textbf {\bibinfo {volume} {9}},\ \bibinfo {pages} {2103013} (\bibinfo {year} {2022})}\BibitemShut {NoStop}%
\bibitem [{\citenamefont {Swift}\ \emph {et~al.}(2024)\citenamefont {Swift}, \citenamefont {Efros},\ and\ \citenamefont {Erwin}}]{swift_controlling_2024}%
  \BibitemOpen
  \bibfield  {author} {\bibinfo {author} {\bibfnamefont {M.~W.}\ \bibnamefont {Swift}}, \bibinfo {author} {\bibfnamefont {A.~L.}\ \bibnamefont {Efros}},\ and\ \bibinfo {author} {\bibfnamefont {S.~C.}\ \bibnamefont {Erwin}},\ }\bibfield  {title} {\bibinfo {title} {Controlling light emission from semiconductor nanoplatelets using surface chemistry},\ }\href {https://doi.org/10.1038/s41467-024-51842-4} {\bibfield  {journal} {\bibinfo  {journal} {Nat Commun}\ }\textbf {\bibinfo {volume} {15}},\ \bibinfo {pages} {7737} (\bibinfo {year} {2024})}\BibitemShut {NoStop}%
\bibitem [{\citenamefont {Rowland}\ \emph {et~al.}(2015)\citenamefont {Rowland}, \citenamefont {Fedin}, \citenamefont {Zhang}, \citenamefont {Gray}, \citenamefont {Govorov}, \citenamefont {Talapin},\ and\ \citenamefont {Schaller}}]{rowland_picosecond_2015}%
  \BibitemOpen
  \bibfield  {author} {\bibinfo {author} {\bibfnamefont {C.~E.}\ \bibnamefont {Rowland}}, \bibinfo {author} {\bibfnamefont {I.}~\bibnamefont {Fedin}}, \bibinfo {author} {\bibfnamefont {H.}~\bibnamefont {Zhang}}, \bibinfo {author} {\bibfnamefont {S.~K.}\ \bibnamefont {Gray}}, \bibinfo {author} {\bibfnamefont {A.~O.}\ \bibnamefont {Govorov}}, \bibinfo {author} {\bibfnamefont {D.~V.}\ \bibnamefont {Talapin}},\ and\ \bibinfo {author} {\bibfnamefont {R.~D.}\ \bibnamefont {Schaller}},\ }\bibfield  {title} {\bibinfo {title} {Picosecond energy transfer and multiexciton transfer outpaces {Auger} recombination in binary {CdSe} nanoplatelet solids},\ }\href {https://doi.org/10.1038/nmat4231} {\bibfield  {journal} {\bibinfo  {journal} {Nature Mater}\ }\textbf {\bibinfo {volume} {14}},\ \bibinfo {pages} {484} (\bibinfo {year} {2015})}\BibitemShut {NoStop}%
\bibitem [{\citenamefont {Taghipour}\ \emph {et~al.}(2018)\citenamefont {Taghipour}, \citenamefont {Hernandez~Martinez}, \citenamefont {Ozden}, \citenamefont {Olutas}, \citenamefont {Dede}, \citenamefont {Gungor}, \citenamefont {Erdem}, \citenamefont {Perkgoz},\ and\ \citenamefont {Demir}}]{taghipour_near-unity_2018}%
  \BibitemOpen
  \bibfield  {author} {\bibinfo {author} {\bibfnamefont {N.}~\bibnamefont {Taghipour}}, \bibinfo {author} {\bibfnamefont {P.~L.}\ \bibnamefont {Hernandez~Martinez}}, \bibinfo {author} {\bibfnamefont {A.}~\bibnamefont {Ozden}}, \bibinfo {author} {\bibfnamefont {M.}~\bibnamefont {Olutas}}, \bibinfo {author} {\bibfnamefont {D.}~\bibnamefont {Dede}}, \bibinfo {author} {\bibfnamefont {K.}~\bibnamefont {Gungor}}, \bibinfo {author} {\bibfnamefont {O.}~\bibnamefont {Erdem}}, \bibinfo {author} {\bibfnamefont {N.~K.}\ \bibnamefont {Perkgoz}},\ and\ \bibinfo {author} {\bibfnamefont {H.~V.}\ \bibnamefont {Demir}},\ }\bibfield  {title} {\bibinfo {title} {Near-{Unity} {Efficiency} {Energy} {Transfer} from {Colloidal} {Semiconductor} {Quantum} {Wells} of {CdSe}/{CdS} {Nanoplatelets} to a {Monolayer} of {MoS$_2$}},\ }\href {https://doi.org/10.1021/acsnano.8b04119} {\bibfield  {journal} {\bibinfo  {journal} {ACS Nano}\ }\textbf {\bibinfo {volume} {12}},\ \bibinfo {pages} {8547} (\bibinfo {year} {2018})}\BibitemShut {NoStop}%
\bibitem [{\citenamefont {Erdem}\ \emph {et~al.}(2019)\citenamefont {Erdem}, \citenamefont {Gungor}, \citenamefont {Guzelturk}, \citenamefont {Tanriover}, \citenamefont {Sak}, \citenamefont {Olutas}, \citenamefont {Dede}, \citenamefont {Kelestemur},\ and\ \citenamefont {Demir}}]{erdem_orientation-controlled_2019}%
  \BibitemOpen
  \bibfield  {author} {\bibinfo {author} {\bibfnamefont {O.}~\bibnamefont {Erdem}}, \bibinfo {author} {\bibfnamefont {K.}~\bibnamefont {Gungor}}, \bibinfo {author} {\bibfnamefont {B.}~\bibnamefont {Guzelturk}}, \bibinfo {author} {\bibfnamefont {I.}~\bibnamefont {Tanriover}}, \bibinfo {author} {\bibfnamefont {M.}~\bibnamefont {Sak}}, \bibinfo {author} {\bibfnamefont {M.}~\bibnamefont {Olutas}}, \bibinfo {author} {\bibfnamefont {D.}~\bibnamefont {Dede}}, \bibinfo {author} {\bibfnamefont {Y.}~\bibnamefont {Kelestemur}},\ and\ \bibinfo {author} {\bibfnamefont {H.~V.}\ \bibnamefont {Demir}},\ }\bibfield  {title} {\bibinfo {title} {Orientation-{Controlled} {Nonradiative} {Energy} {Transfer} to {Colloidal} {Nanoplatelets}: {Engineering} {Dipole} {Orientation} {Factor}},\ }\href {https://doi.org/10.1021/acs.nanolett.9b00681} {\bibfield  {journal} {\bibinfo  {journal} {Nano Lett.}\ }\textbf {\bibinfo {volume} {19}},\ \bibinfo {pages} {4297} (\bibinfo {year} {2019})}\BibitemShut {NoStop}%
\bibitem [{\citenamefont {Yu}\ \emph {et~al.}(2003)\citenamefont {Yu}, \citenamefont {Qu}, \citenamefont {Guo},\ and\ \citenamefont {Peng}}]{yu_experimental_2003}%
  \BibitemOpen
  \bibfield  {author} {\bibinfo {author} {\bibfnamefont {W.~W.}\ \bibnamefont {Yu}}, \bibinfo {author} {\bibfnamefont {L.}~\bibnamefont {Qu}}, \bibinfo {author} {\bibfnamefont {W.}~\bibnamefont {Guo}},\ and\ \bibinfo {author} {\bibfnamefont {X.}~\bibnamefont {Peng}},\ }\bibfield  {title} {\bibinfo {title} {Experimental {Determination} of the {Extinction} {Coefficient} of {CdTe}, {CdSe}, and {CdS} {Nanocrystals}},\ }\href {https://doi.org/10.1021/cm034081k} {\bibfield  {journal} {\bibinfo  {journal} {Chem. Mater.}\ }\textbf {\bibinfo {volume} {15}},\ \bibinfo {pages} {2854} (\bibinfo {year} {2003})}\BibitemShut {NoStop}%
\bibitem [{\citenamefont {Yeltik}\ \emph {et~al.}(2015)\citenamefont {Yeltik}, \citenamefont {Delikanli}, \citenamefont {Olutas}, \citenamefont {Kelestemur}, \citenamefont {Guzelturk},\ and\ \citenamefont {Demir}}]{yeltik_experimental_2015}%
  \BibitemOpen
  \bibfield  {author} {\bibinfo {author} {\bibfnamefont {A.}~\bibnamefont {Yeltik}}, \bibinfo {author} {\bibfnamefont {S.}~\bibnamefont {Delikanli}}, \bibinfo {author} {\bibfnamefont {M.}~\bibnamefont {Olutas}}, \bibinfo {author} {\bibfnamefont {Y.}~\bibnamefont {Kelestemur}}, \bibinfo {author} {\bibfnamefont {B.}~\bibnamefont {Guzelturk}},\ and\ \bibinfo {author} {\bibfnamefont {H.~V.}\ \bibnamefont {Demir}},\ }\bibfield  {title} {\bibinfo {title} {Experimental {Determination} of the {Absorption} {Cross}-{Section} and {Molar} {Extinction} {Coefficient} of {Colloidal} {CdSe} {Nanoplatelets}},\ }\href {https://doi.org/10.1021/acs.jpcc.5b09275} {\bibfield  {journal} {\bibinfo  {journal} {J. Phys. Chem. C}\ }\textbf {\bibinfo {volume} {119}},\ \bibinfo {pages} {26768} (\bibinfo {year} {2015})}\BibitemShut {NoStop}%
\bibitem [{\citenamefont {Rod\`a}\ \emph {et~al.}(2022)\citenamefont {Rod\`a}, \citenamefont {Geiregat}, \citenamefont {Di~Giacomo}, \citenamefont {Moreels},\ and\ \citenamefont {Hens}}]{roda_area-independence_2022}%
  \BibitemOpen
  \bibfield  {author} {\bibinfo {author} {\bibfnamefont {C.}~\bibnamefont {Rod\`a}}, \bibinfo {author} {\bibfnamefont {P.}~\bibnamefont {Geiregat}}, \bibinfo {author} {\bibfnamefont {A.}~\bibnamefont {Di~Giacomo}}, \bibinfo {author} {\bibfnamefont {I.}~\bibnamefont {Moreels}},\ and\ \bibinfo {author} {\bibfnamefont {Z.}~\bibnamefont {Hens}},\ }\bibfield  {title} {\bibinfo {title} {Area-{Independence} of the {Biexciton} {Oscillator} {Strength} in {CdSe} {Colloidal} {Nanoplatelets}},\ }\href {https://doi.org/10.1021/acs.nanolett.2c03683} {\bibfield  {journal} {\bibinfo  {journal} {Nano Lett.}\ }\textbf {\bibinfo {volume} {22}},\ \bibinfo {pages} {9537} (\bibinfo {year} {2022})}\BibitemShut {NoStop}%
\bibitem [{\citenamefont {Fox}(2008)}]{fox_optical_2008}%
  \BibitemOpen
  \bibfield  {author} {\bibinfo {author} {\bibfnamefont {M.}~\bibnamefont {Fox}},\ }\href@noop {} {\emph {\bibinfo {title} {Optical properties of solids}}},\ \bibinfo {edition} {repr., with corr}\ ed.,\ \bibinfo {series} {Oxford master series in physics {Condensed} matter physics}\ No.~\bibinfo {number} {3}\ (\bibinfo  {publisher} {Oxford Univ. Press},\ \bibinfo {address} {Oxford},\ \bibinfo {year} {2008})\BibitemShut {NoStop}%
\bibitem [{\citenamefont {Landau}\ and\ \citenamefont {Lifšic}(2007)}]{landau_quantum_2007}%
  \BibitemOpen
  \bibfield  {author} {\bibinfo {author} {\bibfnamefont {L.~D.}\ \bibnamefont {Landau}}\ and\ \bibinfo {author} {\bibfnamefont {E.~M.}\ \bibnamefont {Lifšic}},\ }\href@noop {} {\emph {\bibinfo {title} {Quantum mechanics: non-relativistic theory}}},\ \bibinfo {edition} {3rd}\ ed.,\ \bibinfo {series} {Course of theoretical physics / by {L}. {D}. {Landau} and {E}. {M}. {Lifshitz}}\ No.\ \bibinfo {number} {Vol. 3}\ (\bibinfo  {publisher} {Elsevier [u.a.]},\ \bibinfo {address} {Singapore},\ \bibinfo {year} {2007})\BibitemShut {NoStop}%
\bibitem [{\citenamefont {Efros}\ and\ \citenamefont {Rosen}(2000)}]{efros_electronic_2000}%
  \BibitemOpen
  \bibfield  {author} {\bibinfo {author} {\bibfnamefont {A.~L.}\ \bibnamefont {Efros}}\ and\ \bibinfo {author} {\bibfnamefont {M.}~\bibnamefont {Rosen}},\ }\bibfield  {title} {\bibinfo {title} {The {Electronic} {Structure} of {Semiconductor} {Nanocrystals}},\ }\href {https://doi.org/10.1146/annurev.matsci.30.1.475} {\bibfield  {journal} {\bibinfo  {journal} {Annu. Rev. Mater. Sci.}\ }\textbf {\bibinfo {volume} {30}},\ \bibinfo {pages} {475} (\bibinfo {year} {2000})}\BibitemShut {NoStop}%
\bibitem [{\citenamefont {Bertrand}\ \emph {et~al.}(2016)\citenamefont {Bertrand}, \citenamefont {Polovitsyn}, \citenamefont {Christodoulou}, \citenamefont {Khan},\ and\ \citenamefont {Moreels}}]{bertrand_shape_2016}%
  \BibitemOpen
  \bibfield  {author} {\bibinfo {author} {\bibfnamefont {G.~H.~V.}\ \bibnamefont {Bertrand}}, \bibinfo {author} {\bibfnamefont {A.}~\bibnamefont {Polovitsyn}}, \bibinfo {author} {\bibfnamefont {S.}~\bibnamefont {Christodoulou}}, \bibinfo {author} {\bibfnamefont {A.~H.}\ \bibnamefont {Khan}},\ and\ \bibinfo {author} {\bibfnamefont {I.}~\bibnamefont {Moreels}},\ }\bibfield  {title} {\bibinfo {title} {Shape control of zincblende {CdSe} nanoplatelets},\ }\href {https://doi.org/10.1039/C6CC05705E} {\bibfield  {journal} {\bibinfo  {journal} {Chem. Commun.}\ }\textbf {\bibinfo {volume} {52}},\ \bibinfo {pages} {11975} (\bibinfo {year} {2016})}\BibitemShut {NoStop}%
\bibitem [{\citenamefont {Dutta}\ \emph {et~al.}(2020)\citenamefont {Dutta}, \citenamefont {Medda}, \citenamefont {Bera}, \citenamefont {Sarkar}, \citenamefont {Sain}, \citenamefont {Kumar},\ and\ \citenamefont {Patra}}]{dutta_hybrid_2020}%
  \BibitemOpen
  \bibfield  {author} {\bibinfo {author} {\bibfnamefont {A.}~\bibnamefont {Dutta}}, \bibinfo {author} {\bibfnamefont {A.}~\bibnamefont {Medda}}, \bibinfo {author} {\bibfnamefont {R.}~\bibnamefont {Bera}}, \bibinfo {author} {\bibfnamefont {K.}~\bibnamefont {Sarkar}}, \bibinfo {author} {\bibfnamefont {S.}~\bibnamefont {Sain}}, \bibinfo {author} {\bibfnamefont {P.}~\bibnamefont {Kumar}},\ and\ \bibinfo {author} {\bibfnamefont {A.}~\bibnamefont {Patra}},\ }\bibfield  {title} {\bibinfo {title} {Hybrid {Nanostructures} of {2D} {CdSe} {Nanoplatelets} for {High}-{Performance} {Photodetector} {Using} {Charge} {Transfer} {Process}},\ }\href {https://doi.org/10.1021/acsanm.0c00728} {\bibfield  {journal} {\bibinfo  {journal} {ACS Appl. Nano Mater.}\ }\textbf {\bibinfo {volume} {3}},\ \bibinfo {pages} {4717} (\bibinfo {year} {2020})}\BibitemShut {NoStop}%
\bibitem [{\citenamefont {Khan}\ \emph {et~al.}(2019)\citenamefont {Khan}, \citenamefont {Pinchetti}, \citenamefont {Tanghe}, \citenamefont {Dang}, \citenamefont {Martín-García}, \citenamefont {Hens}, \citenamefont {Van~Thourhout}, \citenamefont {Geiregat}, \citenamefont {Brovelli},\ and\ \citenamefont {Moreels}}]{khan_tunable_2019}%
  \BibitemOpen
  \bibfield  {author} {\bibinfo {author} {\bibfnamefont {A.~H.}\ \bibnamefont {Khan}}, \bibinfo {author} {\bibfnamefont {V.}~\bibnamefont {Pinchetti}}, \bibinfo {author} {\bibfnamefont {I.}~\bibnamefont {Tanghe}}, \bibinfo {author} {\bibfnamefont {Z.}~\bibnamefont {Dang}}, \bibinfo {author} {\bibfnamefont {B.}~\bibnamefont {Martín-García}}, \bibinfo {author} {\bibfnamefont {Z.}~\bibnamefont {Hens}}, \bibinfo {author} {\bibfnamefont {D.}~\bibnamefont {Van~Thourhout}}, \bibinfo {author} {\bibfnamefont {P.}~\bibnamefont {Geiregat}}, \bibinfo {author} {\bibfnamefont {S.}~\bibnamefont {Brovelli}},\ and\ \bibinfo {author} {\bibfnamefont {I.}~\bibnamefont {Moreels}},\ }\bibfield  {title} {\bibinfo {title} {Tunable and {Efficient} {Red} to {Near}-{Infrared} {Photoluminescence} by {Synergistic} {Exploitation} of {Core} and {Surface} {Silver} {Doping} of {CdSe} {Nanoplatelets}},\ }\href {https://doi.org/10.1021/acs.chemmater.8b05334} {\bibfield  {journal} {\bibinfo  {journal} {Chem. Mater.}\ }\textbf {\bibinfo {volume} {31}},\ \bibinfo {pages} {1450} (\bibinfo {year} {2019})}\BibitemShut {NoStop}%
\bibitem [{\citenamefont {Morrison}\ \emph {et~al.}(2020)\citenamefont {Morrison}, \citenamefont {Sun}, \citenamefont {Yao}, \citenamefont {Loomis},\ and\ \citenamefont {Buhro}}]{morrison_methods_2020}%
  \BibitemOpen
  \bibfield  {author} {\bibinfo {author} {\bibfnamefont {C.}~\bibnamefont {Morrison}}, \bibinfo {author} {\bibfnamefont {H.}~\bibnamefont {Sun}}, \bibinfo {author} {\bibfnamefont {Y.}~\bibnamefont {Yao}}, \bibinfo {author} {\bibfnamefont {R.~A.}\ \bibnamefont {Loomis}},\ and\ \bibinfo {author} {\bibfnamefont {W.~E.}\ \bibnamefont {Buhro}},\ }\bibfield  {title} {\bibinfo {title} {Methods for the {ICP}-{OES} {Analysis} of {Semiconductor} {Materials}},\ }\href {https://doi.org/10.1021/acs.chemmater.0c00255} {\bibfield  {journal} {\bibinfo  {journal} {Chem. Mater.}\ }\textbf {\bibinfo {volume} {32}},\ \bibinfo {pages} {1760} (\bibinfo {year} {2020})}\BibitemShut {NoStop}%
\bibitem [{\citenamefont {Hieu}\ \emph {et~al.}(2024)\citenamefont {Hieu}, \citenamefont {Hong}, \citenamefont {Hanh},\ and\ \citenamefont {Benazzouz}}]{hieu_temperature-dependent_2024}%
  \BibitemOpen
  \bibfield  {author} {\bibinfo {author} {\bibfnamefont {H.~K.}\ \bibnamefont {Hieu}}, \bibinfo {author} {\bibfnamefont {N.~T.}\ \bibnamefont {Hong}}, \bibinfo {author} {\bibfnamefont {P.~T.~M.}\ \bibnamefont {Hanh}},\ and\ \bibinfo {author} {\bibfnamefont {B.~K.}\ \bibnamefont {Benazzouz}},\ }\bibfield  {title} {\bibinfo {title} {Temperature-{Dependent} {Structural} and {Elastic} {Properties} of {CdS}, {CdSe}, and {CdTe} {Compounds} {Studied} by {Statistical} {Moment} {Method}},\ }\href {https://doi.org/10.59277/RomJPhys.2024.69.603} {\bibfield  {journal} {\bibinfo  {journal} {Rom. J. Phys.}\ }\textbf {\bibinfo {volume} {69}},\ \bibinfo {pages} {603} (\bibinfo {year} {2024})}\BibitemShut {NoStop}%
\bibitem [{\citenamefont {Ninomiya}\ and\ \citenamefont {Adachi}(1995)}]{ninomiya_optical_1995}%
  \BibitemOpen
  \bibfield  {author} {\bibinfo {author} {\bibfnamefont {S.}~\bibnamefont {Ninomiya}}\ and\ \bibinfo {author} {\bibfnamefont {S.}~\bibnamefont {Adachi}},\ }\bibfield  {title} {\bibinfo {title} {Optical properties of wurtzite {CdS}},\ }\href {https://doi.org/10.1063/1.360355} {\bibfield  {journal} {\bibinfo  {journal} {J. Appl. Phys.}\ }\textbf {\bibinfo {volume} {78}},\ \bibinfo {pages} {1183} (\bibinfo {year} {1995})}\BibitemShut {NoStop}%
\bibitem [{\citenamefont {Marple}(1964)}]{marple_refractive_1964}%
  \BibitemOpen
  \bibfield  {author} {\bibinfo {author} {\bibfnamefont {D.~T.~F.}\ \bibnamefont {Marple}},\ }\bibfield  {title} {\bibinfo {title} {Refractive {Index} of {ZnSe}, {ZnTe}, and {CdTe}},\ }\href {https://doi.org/10.1063/1.1713411} {\bibfield  {journal} {\bibinfo  {journal} {J. Appl. Phys.}\ }\textbf {\bibinfo {volume} {35}},\ \bibinfo {pages} {539} (\bibinfo {year} {1964})}\BibitemShut {NoStop}%
\end{thebibliography}
%

\clearpage

\setcounter{page}{1}
\setcounter{section}{0}
\setcounter{equation}{0}
\setcounter{figure}{0}
\setcounter{table}{0}
\renewcommand{\thesection}{S-\Roman{section}}
\renewcommand{\theequation}{S\arabic{equation}}
\renewcommand{\thefigure}{S\arabic{figure}}
\renewcommand{\thetable}{S\arabic{table}}

\title{Supporting Information: \\
Extinction Coefficients of CdSe, CdS, and CdTe Nanoplatelets in Solution: A Practical Tool for Concentration Determination}

\maketitle

\begin{table}
\caption{Material parameters for CdSe, CdS, and CdTe NPLs.  Here $\Delta$ is the spin-orbit splitting band-structure parameter.  Most parameters are from Ref.~\citenum{ithurria_colloidal_2011}. $\epsilon_i$ is taken from the refractive indices~\cite{ninomiya_optical_1995,marple_refractive_1964}.  $v_\text{ph} = \sqrt{G/\rho}$ where $\rho$ is the density and $G$ is the shear modulus from ref.~\citenum{hieu_temperature-dependent_2024}.  $A$, $B$, $C$, and $D$ are the fit parameters for Eq.~\ref{eq:hbar_omega}}
\label{s_tab:material}
\begin{tabular}{llll}
\toprule
 & CdSe & CdS & CdTe \\
\midrule
$\alpha$ & $-0.179$ & $-3.570$ & $-0.130$ \\
$E_\text{p}$ (eV) & $17.5$ & $21.0$ & $21.0$ \\
$E_\text{g}$ (eV) & $1.66$ & $2.40$ & $1.49$ \\
$\Delta$ (eV) & $0.942$ & $0.062$ & $0.600$ \\
$a_\text{cubic}$ (nm) & $0.608$ & $0.582$ & $0.648$ \\
$m_\text{h}$ ($m_0$) & $0.190$ & $0.227$ & $0.110$ \\
$\epsilon_\text{i}$ & $6.10$ & $4.65$ & $8.71$ \\
$\epsilon_\text{o}$ & $2.25$ & $2.25$ & $2.25$ \\
$v_\text{ph}$ (cm/s) & $1.895\times 10^5$ & $2.254\times 10^5$ & $1.614\times 10^5$ \\
$A$ & $1.443$ & $2.138$ & $1.417$ \\
$B$ & $3.207$ & $18.541$ & $1.796$ \\
$C$ & $0.272$ & $-0.017$ & $0.776$ \\
$D$ & $2.633$ & $17.527$ & $1.731$ \\
\bottomrule
\end{tabular}
\end{table}

\begin{table}
\caption{Thickness-dependent parameters for CdSe NPLs}
\label{s_tab:CdSe}
\begin{tabular}{lrrrrrr}
  \toprule
 & $d$ (nm) & $E_\text{ex}$ (eV) & $m_\text{e}$ ($m_0$) & $\mathcal{Q}$ & $a_\text{2D}$ (nm) & $E_\text{b}$ (eV) \\
 \midrule
2.5 ML & $0.608$ & $3.167$ & $0.222$ & $0.395$ & $1.347$ & $-0.375$ \\
3.5 ML & $0.912$ & $2.687$ & $0.197$ & $0.518$ & $1.621$ & $-0.299$ \\
4.5 ML & $1.216$ & $2.424$ & $0.177$ & $0.611$ & $1.845$ & $-0.252$ \\
5.5 ML & $1.520$ & $2.249$ & $0.164$ & $0.681$ & $2.080$ & $-0.220$ \\
6.5 ML & $1.824$ & $2.124$ & $0.154$ & $0.734$ & $2.256$ & $-0.196$ \\
7.5 ML & $2.128$ & $2.031$ & $0.147$ & $0.775$ & $2.420$ & $-0.178$ \\
\bottomrule
\end{tabular}
\end{table}

\begin{table}
\caption{Thickness-dependent parameters for CdS NPLs}
\label{s_tab:CdS}
\begin{tabular}{lrrrrrr}
  \toprule
 & $d$ (nm) & $E_\text{ex}$ (eV) & $m_\text{e}$ ($m_0$) & $\mathcal{Q}$ & $a_\text{2D}$ (nm) & $E_\text{b}$ (eV) \\
 \midrule
2.5 ML & $0.582$ & $3.643$ & $0.463$ & $0.365$ & $0.915$ & $-0.559$ \\
3.5 ML & $0.873$ & $3.243$ & $0.349$ & $0.507$ & $1.115$ & $-0.438$ \\
4.5 ML & $1.164$ & $2.998$ & $0.295$ & $0.610$ & $1.294$ & $-0.366$ \\
5.5 ML & $1.455$ & $2.838$ & $0.265$ & $0.687$ & $1.428$ & $-0.319$ \\
6.5 ML & $1.746$ & $2.727$ & $0.246$ & $0.745$ & $1.550$ & $-0.285$ \\
7.5 ML & $2.037$ & $2.646$ & $0.232$ & $0.789$ & $1.663$ & $-0.259$ \\
\bottomrule
\end{tabular}
\end{table}

\begin{table}
\caption{Thickness-dependent parameters for CdTe NPLs}
\label{s_tab:CdTe}
\begin{tabular}{lrrrrrr}
  \toprule
 & $d$ (nm) & $E_\text{ex}$ (eV) & $m_\text{e}$ ($m_0$) & $\mathcal{Q}$ & $a_\text{2D}$ (nm) & $E_\text{b}$ (eV) \\
 \midrule
2.5 ML & $0.648$ & $3.940$ & $0.201$ & $0.474$ & $2.040$ & $-0.245$ \\
3.5 ML & $0.972$ & $2.901$ & $0.149$ & $0.529$ & $2.552$ & $-0.189$ \\
4.5 ML & $1.296$ & $2.475$ & $0.128$ & $0.596$ & $2.942$ & $-0.158$ \\
5.5 ML & $1.620$ & $2.238$ & $0.116$ & $0.656$ & $3.265$ & $-0.137$ \\
6.5 ML & $1.944$ & $2.086$ & $0.109$ & $0.706$ & $3.600$ & $-0.122$ \\
7.5 ML & $2.268$ & $1.979$ & $0.104$ & $0.747$ & $3.807$ & $-0.111$ \\
\bottomrule
\end{tabular}
\end{table}

\clearpage

\section{Excitons in II-VI compound NPLs}

\begin{figure}
\includegraphics[width=\linewidth]{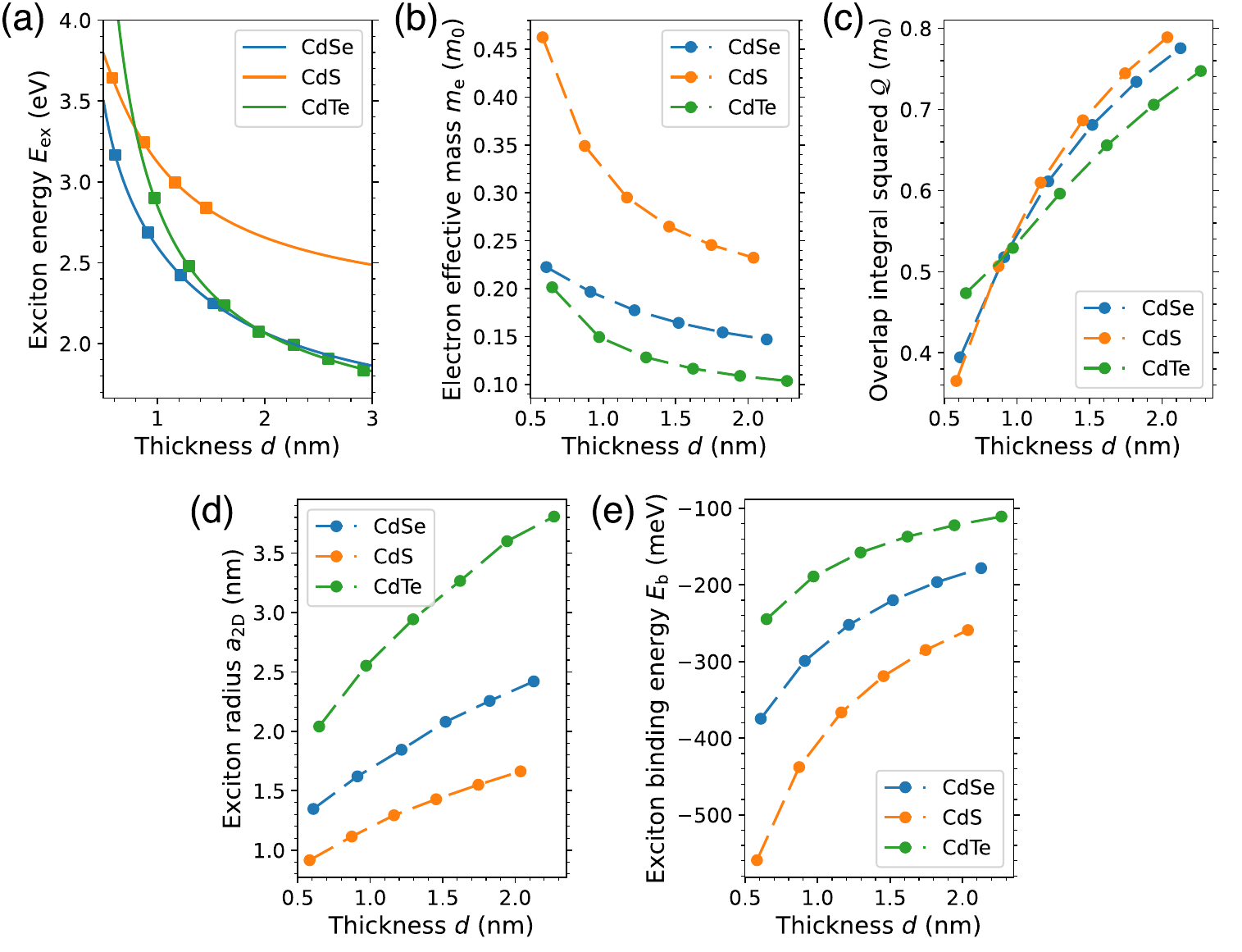}
\caption{Exciton parameters in CdSe, CdS, and CdTe as a function of the NPL thickness $d$.  (a) Exciton emission energy (eq.~\ref{eq:hbar_omega}) (b) Electron effective mass (ref.~\citenum{shornikova_addressing_2018} for CdSe, Eq.~\ref{eq:me_kdotp} for CdS and CdTe).  (c) $\Q$ from Eq.~\ref{eq:Q}.  (d) 2D excitons' radius $a_{\rm 2D}$ and (e) Exciton binding energy $E_\text{b}$ as determined variationally using the ansatz Eq.~\ref{eq:wf}.} 
\label{sfig:params}
\end{figure}

To calculate exciton radius and binding energy we use the ansatz  function
\begin{equation}
    \phi^d_{\rm 2D}(r) = \frac{\sqrt{2}}{a_{\rm 2D}(d)\sqrt{\pi}} \exp\left(-\frac{r}{a_{\rm 2D}(d)}\right)~.
\label{eq:wf}
\end{equation}
The 2D exciton radius $a_{\rm 2D}$ is set to its optimum value, found numerically by minimizing the total energy of $\phi_{\rm 2D}^d$.  The corresponding binding energy is $E_b$.  These quantities are shown as a function of NPL thickness in Figure~\ref{sfig:params}.  

We use a physically-motivated empirical fit for the exciton energy~\cite{swift_dark_2022}:
\begin{equation}
    E_\text{ex}(d) = A + C/d^{2} + \left[B/(D+1/d^2)\right]/d \label{eq:hbar_omega}~.
\end{equation}
With $\hbar\omega$ expressed in meV and $d$ expressed in nm, the parameters fitted to the thickness-corrected data from refs.~\citenum{ithurria_quasi_2008,efros_nanocrystal_2021} may be found in Table~\ref{s_tab:material} and are plotted in Figure~\ref{sfig:params}.

The electron effective mass in CdSe was calculated using an 8-band k$\cdot$p model in ref.~\citenum{shornikova_addressing_2018}.  This calculation was also used in ref.~\citenum{swift_controlling_2024} and was shown to be qualitatively consistent with a simplified expression:
\begin{equation}
    \frac{m_0}{m_\text{e}^*(d)} = \alpha + \frac{E_\text{p}}{3}\left(\frac{2}{E_\text{g}(d)} + \frac{1}{E_\text{g}(d) + \Delta} \right),
    \label{eq:me_kdotp}
\end{equation}
We use the simplified expression for the effective mass in CdS and CdTe.  The results for all three materials are plotted in Figure~\ref{sfig:params}.

\section{Orientation averaging}

If the light enters the NP at an angle  $\theta$ to the surface ($\theta$ varies from 0 at normal incidence to $\pi/2$)  the strength of the light coupling with the exciton dipole is proportional to $\cos^2\theta$. Averaging over all possible  NP orientations results in the following average coupling parameter, $\gamma$:
\begin{equation}
\gamma= {\int_0^{2 \pi}\int_0^{\pi/2} (\cos^2\phi \cos^2\theta + \sin^2\phi) \sin \theta d\theta d\phi\over \int_0^{2 \pi}\int_0^{\pi/2} \sin \theta d\theta d\phi}={2\over 3} ~.
\end{equation}

This result constitutes the average for p-polarized light ($\phi = 0$) and s- polarized light ($\phi = \pi/2$), which separately have the values, $\gamma_p = 1/3$, $\gamma_s = 1$.
  
 This estimate allows us to obtain  the  absorption coefficient of an ensemble of randomly oriented NPs as
 \begin{equation}
 \alpha(\omega)={2\over3}N_{\rm NP} S(\omega)
\label{eq:2}
\end{equation}

\section{Cross-section derivation}
The cross-section, $S(\omega)$,  for any nanoparticle or atom is determined from the relationship;
\begin{equation}
 I(\omega)(c/n)S(\omega)=\hbar\omega  P(\omega)~,
 \label{eq:S_def_SI}
 \end{equation}
where $I(\omega)$ is the energy density  of the light and $P(\omega)$   is  total excitation probability of single particle by the light. This  is  the {\bf definition} of the absorption crossection, the parameter which describes us how much energy is  absorbed by atom or nonocrystals    from the flow of the electromagnetic energy of light. The last energy  is determined by the energy of electromagnetic field  multiply by its speed  in the solid $c/n$.

Let us calculate  the excitation probability  of the  nanoparticles that is  normal to propagation light,  $P(\omega)$. The electric field, $E_x$,   and magnetic field, $H_y$, components  in the  linear polarized light wave that moves in $z$  direction can we written as:
\begin{equation}
H_y=E_x= b\sin\omega(z/c-t)~, ~ H_x= E_y=0~.
\label{eq:16}
\end{equation}
We can rewrite this equation introducing vector potential $\bm A$ and scalar  potential $\phi$, using the relations:
\begin{equation}
{\bm E}=-{1\over c}{\partial {\bm A}\over \partial t} - {\bm \nabla}~ \phi ~, ~ {\bm H}= (\nabla \times {\bm A})
\end{equation}
Defining $\phi=0$ and 
\begin{equation}
A_x= a\cos\omega(z/c-t)~,~ A_y=A_z=0 
\end{equation}
where we have defined $a=bc/\omega$. The energy density $I(\omega)$ of the the light wave is
\begin{equation}
\overline{{E^2+H^2\over 8\pi}}= {a^2\omega^2\over 8\pi c^2}=I(\omega)
\end{equation}

The light interaction with matter is described  by the term: $-i(e/cm_0)({\bm A}\cdot {\bm p})$

The transition probabilities to the continuous spectrum under periodic perturbation in unit time is described by Fermi's golden rule:
\begin{equation}
P_{0,i}={2\pi\over \hbar}\sum_i|F_{i,0}|^2\delta(E_i-\hbar\omega)
\end{equation}
which was obtained from the time dependent matrix element (see \cite{landau_quantum_2007})
\begin{equation}
a_{0,i}=-F_{i,0}{e^{i(\omega_{i0}-\omega)t}-1\over \hbar(\omega_{i0}-\omega)}~,
\end{equation}
where, defining $x$ to be along the projection of the electric field in the NPL plane, we arrive to 
\begin{equation}
F_{i,0}={ea\over 2m_0c}\langle i|\hat{p}_x|0\rangle~, ~{\rm and}~ |F_{i,0}|^2={e^2a^2\over 4m^2_0c^2}|\langle i|\hat{p}_x|0\rangle|^2={2\pi e^2\over m^2_0\omega^2}I(\omega)|\langle i|\hat{p}_x|0\rangle|^2~.
\label{eq:15}
\end{equation}

Now let us estimate the matrix element in CdSe NPLs.  For the heavy hole subband we used Bloch function basis $(\ket{X}\pm i\ket{Y})/\sqrt{2}$  that allows to extract the  main contribution to  matrix element: 
\begin{equation}
    \left|\frac{\bra{X}\pm i\bra{Y}}{\sqrt{2}}\hat{p}_x\ket{0}\right|^2 = \frac{P_{\rm cv}^2}{2}~,
    \label{eq:Bloch_matrix_element}
\end{equation}
where $P_{\rm cv}=\langle S|\hat{p}_x|X\rangle$ is the Kane matrix element.  The full matrix element is
\begin{equation}
|\langle i|\hat{p}_x|0\rangle|^2={P_{\rm cv}^2 \over 2}|\phi_{\rm 2D}^d(0)|^2\left|\int d^2R\Psi_i(\bm R)\right|^2 \left|\int_0^d dz f_\e(z)f_\text{h} (z)\right|^2~.
\label{eq:17}
\end{equation}
Here $\phi_\text{2D}^d$ is the wave-function of the in-plane relative electron-hole motion defined in Eq. \eqref{eq:wf}$, \Psi_i$ is the wave-function envelope for in-plane center-of-mass motion, and $f_\e$ and $f_\text{h}$ are the electron and hole wavefunctions in the $z$ direction.  We define the $z$-direction overlap integral:
\begin{equation}
\Q = \left|\int_0 ^ddz f_\e(z)f_\text{h} (z)\right|^2~,
\end{equation}
which in parabolic band approximation equal to unity: $\Q=1$.  The non-parabolic  character of the electron spectrum  spectrum  leads to admixture of the valence band to the conduction band and reduce the overlap integral. In section~\ref{sec:Q} we calculate $\Q$ and  show its dependence on the NPL thickness $d$.  

Substituting Eq. \eqref{eq:17} to Eq. \eqref{eq:15} we obtain:
\begin{equation}
|F_{x,0}|^2={\pi e^2E_\text{p}\over 2m_0\omega^2}I(\omega)\Q|\phi_{\rm 2D}^d(0)|^2\left|\int d^2R\Psi_i(\bm R)\right|^2~,
\end{equation}
where we introduce Kane energy $E_\text{p} = 2 P_{\rm cv}^2/m_0$.  
The total transition probability can be written:
\begin{equation}
P(\omega) ={\pi^2 e^2E_\text{p}\over \hbar m_0\omega^2}I(\omega)\Q f_T|\phi_{\rm 2D}^d(0)|^2\sum_i \left|\int d^2R\Psi_i(\bm R)\right|^2\delta(E_i-\hbar\omega)~,
\label{eq:P_omega}
\end{equation}
Where we have introduced the factor $f_T$ to take phonon-induced decoherence into account; see Eq.~\eqref{eq:f_T} in the main text. This gives us the absorption cross section $S(\omega)$ defined in Eq.\eqref{eq:S_def_SI}
\begin{eqnarray}
S(\omega)&=&{e^2\over c}{\pi^2 nE_\text{p}\over  m_0\omega}\Q f_T|\phi_{\rm 2D}^d(0)|^2 \sum_i\left|\int d^2R\Psi_i(\bm R)\right|^2\delta(E_i-\hbar\omega)
\end{eqnarray}

Using Eq.~\eqref{eq:wf}, we see that 
\begin{equation}
|\phi_{\rm 2D}^d(0)|^2 = \frac{2}{\pi a_\text{2D}^2}
\end{equation}
\begin{eqnarray}
S(\omega)&=&{e^2\over c}{2\pi nE_\text{p}\over  m_0\omega a_\text{2D}^2}\Q f_T \sum_i\left|\int d^2R\Psi_i(\bm R)\right|^2\delta(E_i-\hbar\omega)~.
\end{eqnarray}
Furthermore we use the fine structure constant $e^2/(\hbar c)\approx 1/137$, introduce $f_T$ discussed in the main text, and substitute $E_\text{ex}=\hbar\omega$ in the denominator, we arrive at Eq.~\ref{eq:S_omega} from the main text:
\begin{eqnarray}
S(\omega)&=&{2\hbar^2\pi nE_\text{p}\over 137 m_0E_\text{ex} a_\text{2D}^2}\Q f_T \sum_i\left|\int d^2R\Psi_i(\bm R)\right|^2\delta(E_i-\hbar\omega)
\end{eqnarray}

\section{Overlap Integral between conduction and valence bands }
\label{sec:Q}

It is often assumed that the $z$-direction overlap integral $\Q = 1$.  However, this neglects the fact that the electron wavefunction has some admixture of the valence band: $f_\e = f_\text{c} + f_\text{v}^1 + f_\text{v}^2$.
The conduction band envelope function is $f_\text{c} = A \cos (\pi z/d) $.  The two valence band envelope components of the electron wavefunction can be obtained from Eq. 4 of  Ref. \cite{efros_electronic_2000}: 
\begin{eqnarray}
f_\text{v}^1 & =& \sqrt\frac{2}{3}\hat{p}_zf_\text{c} \frac{V}{E_g+ \epsilon} = -i\hbar A\sqrt\frac{2}{3}\frac{\pi }{d}\sin\left(\frac{\pi z}{d}\right) \frac{V}{E_g+ \epsilon}\nonumber\\
f_\text{v}^2 & =& -\sqrt\frac{1}{3}\hat{p}_zf_\text{c} \frac{V}{E_g+ \epsilon+\Delta} = -i\hbar A\sqrt\frac{1}{3}\frac{\pi }{d}\sin\left(\frac{\pi z}{d}\right) \frac{V}{E_g+ \epsilon+\Delta} ~.
\end{eqnarray}
Where $V = -i \bra{S}\hat{p}_z\ket{Z}/m_0$, $E_g$ is the band gap, $\epsilon$ is the electron energy, and $\Delta$ is the spin-orbit splitting of the valence band.  The normalization constant $A$ can be obtained from
\begin{equation}
 \int_{-d/2}^{d/2} dz |f_\e(z)|^2  = \int_{-d/2}^{d/2} dz \left(|f_\text{c}(z)|^2+|f_v^2(z)|^2+ |f_v^1(z)|^2\right)  = 1
\end{equation}
Following Ref.~\citenum{swift_controlling_2024}, we assume $E_g + \epsilon = E_\text{ex}(d)$. We can now write: 
\be
A^2 \int_{0}^{d}\left[ {\cos^2 \left( \frac{\pi z}{d}  \right)} + \frac{1}{3} \hbar^2 \frac {\pi^2}{d^2} \sin^2 \left(  \frac{\pi z}{d} \right)   \frac{V^2}{\left(E_\text{ex}(d) + \Delta \right)^2}  + 
\frac{2}{3} \hbar^2 \frac{\pi^2}{d^2} \sin^2 \left(    \frac{\pi z}{d} \right)  \frac{V^2}{\left( E_\text{ex}(d) \right)^2}   \right] dz   =1~,
\ee
resulting in
\be
A^2\frac{d}{2}\left[1+\frac{\hbar^2\pi^2}{6m_0d^2}\left( \frac{E_\text{p}}{\left(E_\text{ex}(d)+\Delta\right)^2 }+\frac{2E_\text{p}}{\left(E_\text{ex}(d)\right)^2} \right) \right] =1
\ee
where we have used the fact that $E_\text{p}=2m_0 V^2$. 
We can now compute the squared overlap integral $\Q$. The valence band components are orthogonal so vanish from the integral.  Therefore,
\bea
{\cal Q}&=&  A^2\left|\int_{-d/2}^{d/2} dz  f_\text{c}(z) f_h(z)\right|^2  =\frac{2A^2}{d}\left|\int_0^d dz  \cos^2\left(\frac{\pi z}{d}\right) \right|^2 = \frac{A^2 d}{2}\nonumber\\
  &=& \left[1+\frac{\hbar^2\pi^2}{6m_0d^2}\left( \frac{E_\text{p}}{\left(E_\text{ex}(d)+\Delta\right)^2 }+\frac{2E_\text{p}}{\left(E_\text{ex}(d)\right)^2} \right) \right]^{-1}
\label{eq:Q}
\end{eqnarray}
which clearly gives us $\Q < 1$.  The results are plotted in Figure~\ref{sfig:params} and tabulated in Tables~\ref{s_tab:CdSe},\ref{s_tab:CdS},\ref{s_tab:CdTe}.

\section{Experimental Details}

\subsection{Synthesis Chemicals}
Cadmium acetate dihydrate (Cd(OAc)$_2$2H$_2$O$\geq$ 98\%), cadmium nitrate tetrahydrate 
(Cd(NO$_3$)$_2$4H$_2$O$\geq $98\%), sodium myristate (C$_{14}$H$_{27}$NaO$_2$$\geq 99$\%), selenium powder (Se $\geq 99.5$ \%), technical grade 1-octadecene (ODE, 90\%), and technical grade oleic acid (OA, 90\%), methyl acetate (CH$_3$COOCH$_3$, 99\%), and 200-proof HPLC-grade ethanol were purchased from Sigma-Aldrich. HPLC-grade methanol and n-hexane were purchased from Fisher Chemical. All chemicals were used as delivered without any further purification. 
\subsection{Preparation of Cadmium Myristate }
Cadmium myristate was synthesized according to a previously reported method.\cite{khan_tunable_2019}  Three grams of cadmium nitrate tetrahydrate were first dissolved in 200 mL of methanol. In a separate beaker, 5 grams of sodium myristate was dissolved in 500 mL of methanol using vigorous stirring and ultrasonication. The cadmium nitrate solution was then added dropwise to the sodium myristate solution and vigorously stirred for 2 hours. The cloudy white suspension was vacuum filtered using a fine-grit borosilicate glass Buchner funnel to obtain the cadmium myristate powder, which was washed thrice with an excess of methanol. The dried precipitate was kept under vacuum overnight on a Schlenk line, followed by an overnight vacuum in the glovebox antechamber, before storing in a N$_2$·-controlled glovebox for future use.  
\subsection{Synthesis of 4.5 Monolayer CdSe Nanoplatelets:} 
The synthesis of core-only 4.5 monolayer nanoplatelets was guided by a previously reported method.\cite{bertrand_shape_2016} Typically, in a three-neck round-bottom flask, 180mg of cadmium myristate, 30mg of selenium powder, and 15mL 1-octadecene were combined.  The mixture was degassed at room temperature, followed by degassing at 120$^o$C for 1 hour.  The flask was then returned to a nitrogen environment and heated to 240$^o$C while stirring at 1000 rpm.  When the temperature reached 210$^o$C, a calculated amount (70mg-120mg) of cadmium acetate dihydrate was swiftly added to the flask.  The mixture was maintained at 240$^o$C for the required time (7-10 minutes), and was then cooled to room temperature using an air gun and water bath. When cooling, 2mL of oleic acid is injected at 160$^o$C, followed by the addition of 15mL of hexanes at room temperature.  The nanoplatelet solution was divided into Falcon tubes and centrifuged at 3000 RPM for 15 minutes.  The supernatant was discarded, and the pellet redispersed in hexane.  The solution was allowed to sit overnight before centrifuging at 6000 RPM for 10 minutes.  The nanoplatelets were stored in dark and airtight conditions, and the pellet was discarded. 
\subsection{Synthesis of 5.5 Monolayer CdSe Nanoplatelets}
A typical synthesis of core-only 5.5 monolayer nanoplatelets proceeds as follows.\cite{dutta_hybrid_2020}  In a 100mL three-neck round-bottom flask, 340mg of cadmium myristate and 14mL 1-octadecene were combined. The mixture was degassed at room temperature for 1 hour.  The flask was then returned to a nitrogen environment and heated to 250$^o$C while stirring at 1000rpm.  At 250$^o$C, 24mg of selenium powder dispersed in 1mL of 1-octadecene was injected.  The temperature drops to 245-248$^o$C depending on the experimental conditions. After 110-120 seconds, at 250$^o$C, the required amount (140mg-240mg) of cadmium acetate dihydrate is swiftly added. The mixture was maintained at 250$^o$C for the necessary time (6-10 minutes), followed by 2mL of oleic acid injection at 250$^o$C.  The reaction was quenched by cooling to room temperature with an air gun and a water bath. At room temperature, 15mL of hexanes and 10mL of ethanol were directly added to the solution and divided into four Falcon tubes.  The nanoplatelets were precipitated by centrifugation at 4000rpm for 15 minutes. The supernatant was discarded, and the pellet was redispersed in hexane.  The solution was allowed to sit for 1 hour before it was centrifuged at 4000 rpm for 15 minutes.  The nanoplatelet solution was stored in dark and airtight conditions for future use, and the pellet was discarded. 
\subsection{Optical Characterization}
A Shimadzu UV-1800 spectrophotometer was used to measure the absorption spectra of purified NPL samples in hexane, quantify them for elemental analysis, and to calculate absorption coefficients. An Edinburgh Instruments FLS 1000 photoluminescence spectrometer was used to measure emission spectra and confirm sample purity.   
\subsection{Inductively coupled plasma optical emission spectroscopy (ICP-OES)} 
The elemental compositions of NPLs were determined using Agilent Technologies 5100 ICP-OES equipped with an SPS 3 autosampler. Synchronous vertical dual view (SVDV) was used to measure 4 replicates of Cd and Se intensities at 214.439 nm and 196.026 nm, respectively. Nanoplatelet samples were prepared for ICP-OES analysis following a published protocol for digesting semiconductor materials.\cite{morrison_methods_2020} Aliquots of the clean CdSe NPL stock solutions were precipitated and washed with isopropanol and methanol (1:1 by volume) in conical centrifuge tubes, centrifuged to form a pellet, and dried overnight. The pellets were first treated with 30\% H$_2$O$_2$ (0.5 mL) in a tightly capped tube for 5 minutes. Then 69-70\% nitric acid (0.5 mL, J.T. Baker Instra-Analyzed ACS) was added to the tube and digested for 15 minutes or until no solid residue was observed. Finally, the digested NPL solutions were diluted with DI water to reach approximately 1.75\% nitric acid. Calibration standards were prepared by serial dilution of commercial analytical stock solutions containing Cd (1000 ug/mL, Inorganic Ventures) and Se (1000 ug/mL, Spex CertiPrep) with 2
\subsection{Transmission Electron Microscopy (TEM) and Size Analysis} 
\begin{figure}
  \includegraphics[width=\textwidth]{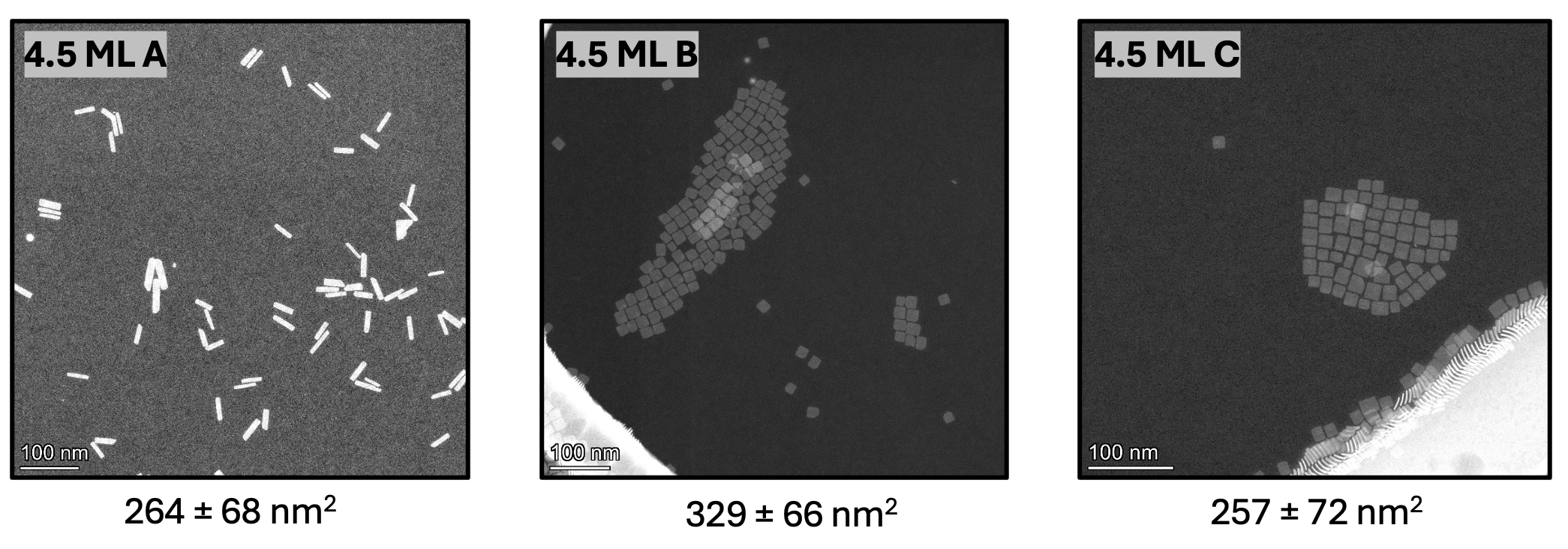}
  \caption{TEM images of 4.5 ML thick CdSe NPLs.}
  \label{fig:TEM4}
\end{figure}

\begin{figure}
  \includegraphics[width=\textwidth]{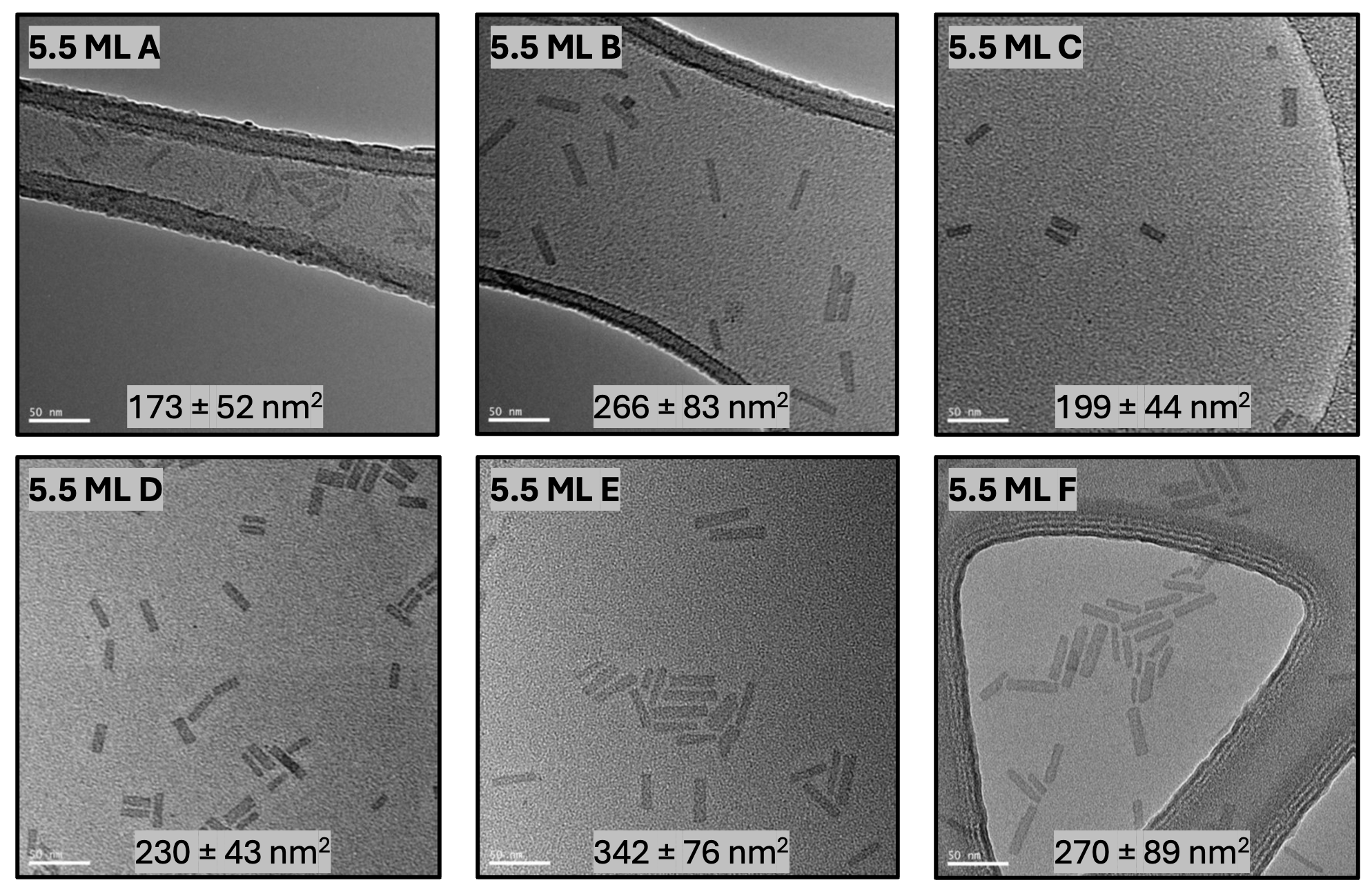}
  \caption{TEM images of 5.5 ML thick CdSe NPLs.}
  \label{fig:TEM5}
\end{figure}

Dilute solutions of 4.5ML   and 5.5 CdSe nanoplatelet samples in hexane were drop cast onto ultrathin carbon film on lacey carbon support film on a 300-mesh gold grid from Ted Pella, Inc. The 4.5 monolayer CdSe NPLs were imaged on a Thermo Scientific Talos F200C TEM using STEM mode at the US Naval Research Lab. 100x-200x solutions of 5.5ML nanoplatelet samples were drop cast onto ultrathin carbon film supported by a lacey carbon film on a 400-mesh copper grid from Ted Pella, Inc. Transmission electron microscopy (TEM) micrographs were collected using a FEI Tecnai F20 field emission microscope with a 200 kV accelerating voltage, maintained by the University of Rochester Integrated Nano-systems Center. Nanoplatelet size statistics were recorded using the National Institutes of Health offered ImageJ software. The TEM images of 4.5 and 5.5 CdSe NPLs are shown in Figs. \ref{fig:TEM4} and \ref{fig:TEM5} correspondingly.

\end{document}